\documentclass{emulateapj}

\usepackage{graphicx}
\usepackage{epstopdf}

\def\m{$\mu$m}
\def\ab{$\sim$}

\def\ha{H$\alpha$}
\def\hb{H$\beta$}
\def\han{H$\alpha$ } 
\def\oiii{[O{\sc  iii}]}

\def\ab{$\sim$}
\def\deg{\ifmmode ^{\circ}
         \else $^{\circ}$\fi} 

\shorttitle{Predicting Future Space Near-IR Grism Surveys}
\shortauthors{Colbert et al.}

\begin{document}
\title{Predicting Future Space Near-IR Grism Surveys using the WFC3
  Infrared Spectroscopic Parallels Survey}
\author{James W. Colbert\altaffilmark{1}, Harry Teplitz\altaffilmark{2}, Hakim Atek\altaffilmark{3,1}, Andrew  Bunker\altaffilmark{4}, 
Marc Rafelski\altaffilmark{2}, Nathaniel Ross\altaffilmark{5}, Claudia  Scarlata\altaffilmark{6},
Alejandro G. Bedregal\altaffilmark{6,7}, Alberto Dominguez\altaffilmark{8}, Alan Dressler\altaffilmark{9}, Alaina
Henry\altaffilmark{10}, Matt Malkan\altaffilmark{5}, Crystal L. Martin\altaffilmark{11}, Dan Masters\altaffilmark{8,9},
Patrick McCarthy\altaffilmark{9}, Brian Siana\altaffilmark{8}}

\altaffiltext{1}{Spitzer Science Center, California Institute of
  Technology, Pasadena, CA 91125}

\altaffiltext{2}{Infrared Processing and Analysis Center, Caltech,
  Pasadena, CA 91125, USA}

\altaffiltext{3}{Laboratoire d'astrophysique, \'Ecole Polytechnique F\'ed\'erale de Lausanne, Observatoire de Sauverny, 1290 Versoix, Switzerland}

\altaffiltext{4}{Department of Physics, University of Oxford, Denys
  Wilkinson Building, Keble Road, Oxford OX1 3RH, UK}

\altaffiltext{5}{Department of Physics and Astronomy, University of
  California, Los Angeles, CA, USA}

\altaffiltext{6}{Minnesota Institute for Astrophysics, University of Minnesota, Minneapolis, MN 55455, USA}

\altaffiltext{7}{Department of Physics and Astronomy, Tufts University, Medford, MA 02155, USA}

\altaffiltext{8}{Department of Physics and Astronomy, University of
California Riverside, Riverside, CA 92521, USA}

\altaffiltext{9}{Observatories of the Carnegie Institution for Science, Pasadena, CA 91101, USA}

\altaffiltext{10}{Astrophysics Science Division, Goddard Space Flight Center, Code 665, Greenbelt, MD 20771}

\altaffiltext{11}{Department of Physics, University of California, Santa Barbara, CA
93106, USA}

\begin{abstract}

We present near-infrared emission line counts and luminosity functions
from the {\it HST} WFC3 Infrared Spectroscopic Parallels
(WISP) program for 29 fields (0.037 deg$^2$) observed using both the G102 and G141
grism. Altogether we identify 1048 emission line galaxies with observed
equivalent widths greater than 40 \AA , 467 of
which have multiple detected emission lines. We use
simulations to correct for significant ($>$20\%) incompleteness
introduced in part by the non-dithered, non-rotated nature of the grism
parallels.
The WISP survey
is sensitive to fainter flux levels (3-5$\times$10$^{-17}$ ergs s$^{-1}$
cm$^{-2}$) than the future space near-infrared grism missions aimed
at baryonic acoustic oscillation cosmology (1-4 $\times$10$^{-16}$ ergs s$^{-1}$
cm$^{-2}$), allowing us to probe the fainter emission line galaxies
that the shallower future surveys may miss. 
Cumulative number counts of 0.7$<$$z$$<$1.5 galaxies 
reach 10,000 deg$^{-2}$ above an \han flux of
2$\times$10$^{-16}$ ergs s$^{-1}$
cm$^{-2}$. \ha -emitting galaxies with comparable [O{\sc
  iii}] flux are roughly 5 times less common than galaxies with just
\han emission at those flux
levels. Galaxies with low \ha /[O{\sc iii}] ratios are very
rare at the brighter fluxes that future near-infrared grism surveys will
probe; our survey finds no galaxies with \ha /[O{\sc iii}] $<$ 0.95 that have \han
flux greater than 3$\times$10$^{-16}$ ergs s$^{-1}$
cm$^{-2}$. Our \han luminosity function contains a comparable
number density of faint line emitters to that found by the NICMOS near-infrared grism
surveys, but significantly fewer (factors of 3-4 less) high luminosity emitters.
We also find that our high redshift ($z$=0.9-1.5)
counts are in agreement with the high redshift
($z$=1.47) narrow band \han survey of HiZELS (Sobral et al. 2013),
while our lower redshift luminosity function ($z$=0.3-0.9) falls
slightly below their $z$=0.84 result. The evolution in both the \han luminosity
function from $z$=0.3--1.5 and the [O{\sc iii}] luminosity function from
$z$=0.7--2.3 is almost entirely in the L$_{\star }$ parameter, which steadily
increases with redshift over those ranges.

\end{abstract}


\section{Introduction}

The majority of star formation, supermassive black hole accretion, and
galaxy assembly in the universe likely occurred over the epoch of
$z$=0.7-2 \citep[e.g.,][]{mcl06,per08,van10,mag11,muz13,ilb13}. A careful census of the spatial distribution of
galaxies in this redshift range is critical for measuring the
large-scale clustering of galaxies that results from
baryonic acoustic oscillations (BAO), which will enable us to probe the
expansion history of the universe and address the equation state of dark
energy \citep{eis98,col05,eis05,wei12}. Large near-infrared
spectroscopic surveys are required to study this redshift regime, as
the most luminous nebular emission lines move out of the optical at
these redshifts.

Ground-based searches for emission lines from faint high-redshift
galaxies are severely impacted by the bright near-IR airglow, which
effectively eliminates the possiblity of slitless grism spectroscopy.
Most near-infrared ground spectroscopy is done a single object at a time, 
requiring some form of pre-selection that generally biases against
discovery of the highest equivalent width sources.
Multi-object near-infrared spectrographs on
the world's largest telescopes (i.e. MOSFIRE, FLAMINGOS-2, MOIRCS,
EMIR, KMOS, etc.) allow for the study of a much greater number of 
galaxies, but are still subject
to the limitations of the available atmospheric transmission
windows. More importantly, multi-object spectroscopy generally 
requires target pre-selection based on broad band photometry, 
creating a bias against discovery of emission line galaxies with the 
most extreme equivalent widths. 

An alternative method to spectroscopy for identifying large numbers of z$>$0.7 emission line
galaxies are wide-field narrow band searches targeting the transmission
windows in our atmosphere, e.g. HiZELS \citep{gea08,sob09,sob12,sob13} and
the NEWFIRM H$\alpha$ Survey \citep{ly11}. These surveys
have the advantage of very high sensitivity to emission lines
(\ab 1$\times$10$^{-17}$ ergs s$^{-1}$), but suffer from an inability
to map outside their narrow redshift ranges, making them unfeasible
for BAO studies. In addition there is the danger of significant contamination
from emission lines at different redshifts, a well known issue from
high redshift Ly$\alpha$ emission line searches
\citep{mar08,hen12}. For instance, the typical \ha /[O{\sc iii}]
5007 \AA\ ratio decreases with redshift
\citep{ly07,van11,dom13}. A
carefully constructed experiment, with multiple photometric bands
along with additional narrow bands targeting another expected line at
the same redshift, can increase the narrow band reliability
\citep[e.g.][]{lee12,sob12}, although surveys dependent on continuum
detections will still miss the lowest
mass galaxies. 

Of course the best way to avoid the limitations of the atmosphere
is to observe in space. Two future space missions, ESA's {\it Euclid}
\citep{lau12,lau11} and
NASA's {\it WFIRST} \citep{gre12,dre12}, both utilize large area, near-infrared grism
surveys to investigate BAO \citep{gla05}, as well as galaxy evolution and star
formation history. In addition to identifying the redshift of galaxies, near-infrared
spectroscopy also allows access to the wealth of stellar evolution diagnostic
features available in the optical: \ha , one of the most reliable indicators
of star formation rate, the Balmer decrement, to determine extinction,
as well as multiple metallicity indicators. These emission line surveys will cover thousands of 
square degrees of continuous redshift space, unbiased by the
underlying continuum luminosity of the galaxy. However, these will not be the first
near-infrared grism observations done from space, as the {\it Hubble 
Space Telescope} has had two instruments with near-infrared grisms.

The Near IR Camera and Multi-Object Spectrometer
\citep[NICMOS,][]{tho98} G141 
grism pure-parallels program \citep{mcc99,yan99,shi09}, surveyed \ab 170 arcmin$^2$, identifying
113 emitters and measuring the \ha -luminosity function from 0.7$< z
<$1.9. The NICMOS data had a tiny ($<$
1 arcmin$^2$) field of view, low spectral resolution (R\ab 100), and relatively
poor detector sensitivity. The Wide Field Camera 3 \citep[WFC3;][]{mac10} instrument is an improvement in
all three areas, providing a 20-fold survey efficiency gain for the
presently on-going WFC3 Infrared Spectroscopic Parallels program
(WISP; Atek et. al 2010, 2011). While much smaller in area, the WISP survey
is much deeper than the future space near-infrared
grism surveys. There is presently no better laboratory for predicting
what these future missions can expect. 

In this paper we discuss a sample of the WISP emission line objects identified
from 0.85 -- 1.65 $\mu$m. We discuss the survey, details of the emission line
extraction, completeness corrections, and estimates of redshift
accuracy. We then present the \han and [O{\sc iii}] line-emitter number counts, H$\alpha$/[O{\sc
  iii}] ratios, and luminosity functions. 
Finally, 
we present our final summary and conclusions.
We assume an $\Omega _{M}$=0.3, $\Omega _{\Lambda }$=0.7 universe with
H$_o$=70 km s$^{-1}$ Mpc$^{-1}$. All magnitudes are in the AB system. 

\section{WISP Survey}

The WISP survey (PI Malkan, GO-11696, GO-12283, GO-12568, \& GO-12902) consists 
of HST WFC3 \citep{kim08} grism observations in uncorrelated high-latitude fields 
obtained in parallel mode while prime observations are being
obtained with the Cosmic Origins Spectrograph \citep[COS;][]{ost11}  or the Space Telescope
Imager and Spectrograph \citep[STIS;][]{woo98}. 
The parallel data include both grism spectroscopy and associated near-IR and optical
imaging. Depending on the length of the parallel opportunity,
the WISP survey either acquires spectroscopy with just the G141
(1.2--1.7 $\mu$m, R\ab 130) grism
or, in the case of longer opportunities, a combination of G141 and the
G102 (0.8--1.2 $\mu$m, R\ab 210) grism with roughly 2-3 times more integration
time spent on the higher resolution G102 grism. At a plate scale of
0.13 arcsec pixel$^{-1}$, the total field of view for each observation
is 123$\arcsec$$\times$136$\arcsec$. We note that one can
only achieve the full grism wavelength resolution for compact sources, as
any spatial extension of the object will also broaden the
features within the spectrum. See \cite{ate10}
for further discussion of the program. 

For this paper we present data from 29 separate fields where we
have both G102 and G141 grism spectroscopy, covering a total of
135 arcmin$^2$ (0.037 deg$^2$) over 159 orbits. These fields, along with their
integration times, are presented in Table \ref{observations}. 
While other large G141-only grism surveys have been
approved by HST \citep[e.g., 3D-HST and AGHAST;][]{bra12,wei12}, no other program comes close to
surveying the area that WISP does with such an extended spectral coverage (0.85--1.65 $\mu$m).

\begin{deluxetable*}{lcccccc} 
\tablecolumns{7} 
\tablewidth{0pc} 
\tablecaption{Summary of WISP Field Observations}
\tablehead{ 
\colhead{Field} & \colhead{RA} & \colhead{Dec} & \colhead{G102} &
\colhead{F110W} & \colhead{G141} & \colhead{F160W} \\
& \colhead{[HMS]} & \colhead{[DMS]} & \colhead{[sec]} &
\colhead{[sec]} & \colhead{[sec]} & \colhead{[sec]} }
\startdata
Par17 & 02:13:38.11 & +12:54:59.3 & 3409 & 534 & 3409 & 559\tablenotemark{a} \\
Par55 & 12:20:54.68 & -02:04:46.0 & 6415 & 909 & 2809 & 484\tablenotemark{a} \\
Par62 & 13:01:16.20 & -00:00:20.2 & 4712 & 734 & 2006 & 396 \\
Par64 & 14:37:29.04 & -01:49:49.5 & 5918 & 1112 & 2306 & 456 \\
Par68 & 23:33:33.04 & +39:21:20.5 & 7721 & 1215 & 3009 & 534 \\
Par69 & 15:24:07.75 & +09:54:53.9 & 5721 & 1087 & 2309 & 431 \\
Par73 & 14:05:12.86 & +46:59:19.9 & 6118 & 1034 & 2509 & 456 \\
Par74 & 09:10:48.14 & +10:17:20.3 & 5918 & 1065 & 2306 & 431 \\
Par76 & 13:27:22.17 & +44:30:39.3 & 5515 & 887 & 2006 & 406 \\
Par78 & 23:28 34.06 & +05:10:28.3 & 5318 & 887 & 2106 & 406 \\
Par79 & 01:10:08.96 & -02:25:16.2 & 7521 & 1187 & 2809 & 534 \\
Par81 & 01:10:09.12 & -02:22:17.1 & 7521 & 1187 & 2809 & 534 \\
Par87 & 09:46:46.39 & +47:14:58.2 & 4915 & 912 & 1906 & 406 \\
Par94 & 22:05:26.66 & -00:17:48.5 & 9024 & 1624 & 3309 & 534 \\
Par96 & 02:09:24.40 & -04:43:41.6 & 28081 & 4295 & 11430 & 1765 \\
Par97 & 01:10:06.30 & -02:23:44.7 & 5515 & 859 & 2109 & 406 \\
Par114 & 10:40:58.09 & +06:07:31.0 & 7221 & 1137 & 2909 & 456 \\
Par115 & 11:18:55.08 & +02:17:09.6 & 5215 & 912 & 2106 & 381 \\
Par120 & 13:56:51.50 & +17:02:33.9 & 4512 & 837 & 1806 & 381 \\
Par124 & 18:32:28.28 & +53:44:50.9 & 4618 & 759 & 1906 & 406 \\
Par129 & 11:02:18.72 & +20:52:07.8 & 4712 & 762 & 2206 & 456 \\
Par131 & 10:48:22.94 & +13:03:50.5 & 13039 & 2171 & 5215 & 884 \\
Par132 & 11:26:19.80 & -01:43:22.1 & 4315 & 634 & 1806 & 356 \\
Par135 & 11:22:24.01 & +57:50:58.9 & 4712 & 862 & 1906 & 406 \\
Par136 & 12:26:28.84 & +05:23:02.9 & 18857 & 3036 & 7318 & 1137 \\
Par143 & 14:02:22.01 & +09:45:51.7 & 10133 & 1568 & 4012 & 759 \\
Par146 & 02:12:27.60 & -07:32:20.7 & 4212 & 887 & 1706 & 381 \\
Par147 & 23:58:19.51 & -10:15:04.6 & 5418 & 962 & 2106 & 406 \\
Par167 & 01:41:24.18 & +13:37:34.1 & 4315 & 659 & 1806 & 356 
\enddata
\tablenotetext{a}{Used F140W as the direct image for identifying
  counterparts in the G141 grism.}
\label{observations}
\end{deluxetable*}

The inclusion of the G102 grism doubles the wavelength and
redshift range surveyed. More importantly, it provides both critical checks of
the assumed redshift and multiple line ratios that can not be obtained
using G141 alone. [O{\sc iii}] and \han only fall together on
the G141 grism over a very narrow redshift range ($z$=1.3--1.55), 
otherwise in the vast majority of cases
G141-only observations only discover single emission lines. 
[O{\sc iii}] lines of comparable or greater strength than \han are common
\citep{hu09,ate10,dom13}, making a catastrophic
misidentification likely. With both G102 and
G141, WISP is sensitive to \ha ,
[O{\sc iii}], \hb , and [O{\sc ii}] over a wide range of
redshifts, and for $z$=1.3--1.55, we are sensitive to all those lines 
simultaneously.

\section{Emission Line Extraction}

All the data were processed with the WFC3 pipeline CALWF3 (ver. 2.1)
to correct for bias, dark, flat-field, and gain variations. 
Then, the slitless extraction package aXe 2.0 \citep{kum09} is
used for the spectral extraction. A complete description of the data
reduction steps is presented in \cite{ate10}.

To locate and identify all the emission lines within our spectra we applied an
automatic line finding algorithm. We first fit a three-segment cubic
spline to the continuum of our one-dimensional aXe-extracted spectra,
using outlier rejection to avoid fitting any emission lines. We then subtract 
the estimated continuum from the spectrum and divide the continuum-subtracted spectrum by the aXe estimate of the flux uncertainties 
for each pixel. This produces a signal-to-noise spectrum. We look for groupings of three or more contiguous pixels 
with an excess above the continuum with signal-to-noise in each pixel greater than $\sqrt{3}$. For an unresolved object, the emission lines 
can be as narrow as two pixels and so we adopt a more-stringent two-pixel criterion of two contiguous pixels with a signal-to-noise ratio 
of greater than $\sqrt{5}$ in each. 
To validate this method of
finding candidate emission lines, we compared several samples of emission
lines found automatically to those found by an intensive, spectra by
spectra, visual inspection and found the automated method did not miss
any significant fraction of lines ($<$5\%). 

The parallel nature of the WISP observations presents several
challenges to emission line extraction. Foremost is the lack of dithering, which greatly
complicates the mitigation of cosmic rays, hot or warm pixels, and
other artifacts. As a practical matter, this means that the automated
line identification process has a very high contamination rate from
false and/or spurious sources. 

To address the false detections we require the visual
inspection of every candidate emission line by two team members, done
independently and without consultation.  After this initial inspection,
we compare the emission line lists from the two reviewers and send
each reviewer back a list of discrepancies, allowing them to make a
second, more careful examination of any emission lines for which there
is a disagreement. This exercise provides a good
final agreement for most emission lines, but we find that for roughly
5\% of the emission lines, consensus can not be reached. These
uncertain lines are removed from the final sample, but are accounted for
in our completeness corrections (see Section \ref{completeness}).
 
\begin{figure*}
\begin{center}
\includegraphics[angle=90,width=7in]{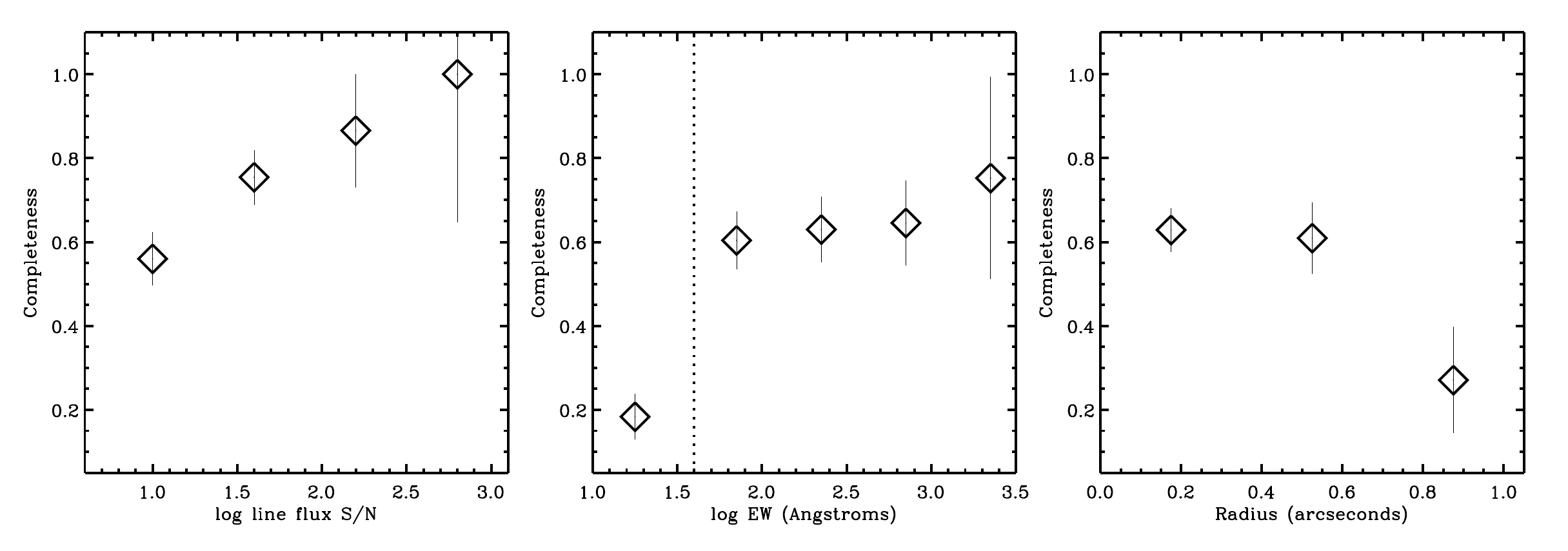}
\caption{Completeness of extracted emission lines as a function of:
 {\it  left)} line flux signal-to-noise, {\it middle)} equivalent width, and
  {\it right)} radius in arcseconds. The dotted line in the middle
  equivalent width plot represents the chosen equivalent width cut.} 
\end{center}
\end{figure*}

The visual inspection consists of examining both the extracted
one dimensional and the original two dimensional spectra, as well as the continuum
fits and measurements of signal-to-noise. There are only four
conditions for which we exclude a candidate line:

1) If the line is clearly an artifact, either a cosmic ray, a zero
order image from a nearby bright source, a diffraction spike, ghost,
or some other artifact clearly not associated with the spectrum in
question.

2) If the continuum fit under the location of the emission line is
highly inaccurate, artificially increasing the significance of the emission
line. Many of the spectra contain breaks, both real and caused by
nearby contamination from overlapping sources, which often lead to continuum errors of this
sort.

3) If the local noise variations are larger than the flux
uncertainties used by the automated line identification software, measuring lines at a higher
significance than local noise conditions actually warranted. While the
WFC3 detector and grism are fairly uniform and well-behaved, we find
variations across the detector and locations where what appear to be
non-Poissonian sources of noise exist. In these cases the automated
recovery pipeline was likely to find every noise peak and call it a
source.

4) If the contamination from nearby overlapping spectra is so great that we
can not determine what source is producing the emission line.
However, in most cases even severe overlapping source contamination can be untangled. We use
multiple lines spread across the G102 and G141 grisms (only the real
source produces the correct wavelength ratios in those cases),
small pixel offsets from source center, and discrepancies between the 
size of the object and the spatial extent of the emission line as
evidence for assigning the confused emission line to its proper
source. Only if the source remains ambiguous after all analysis 
is it excluded from the final sample.

\begin{figure*}
\begin{center}
\includegraphics[angle=90,width=7in]{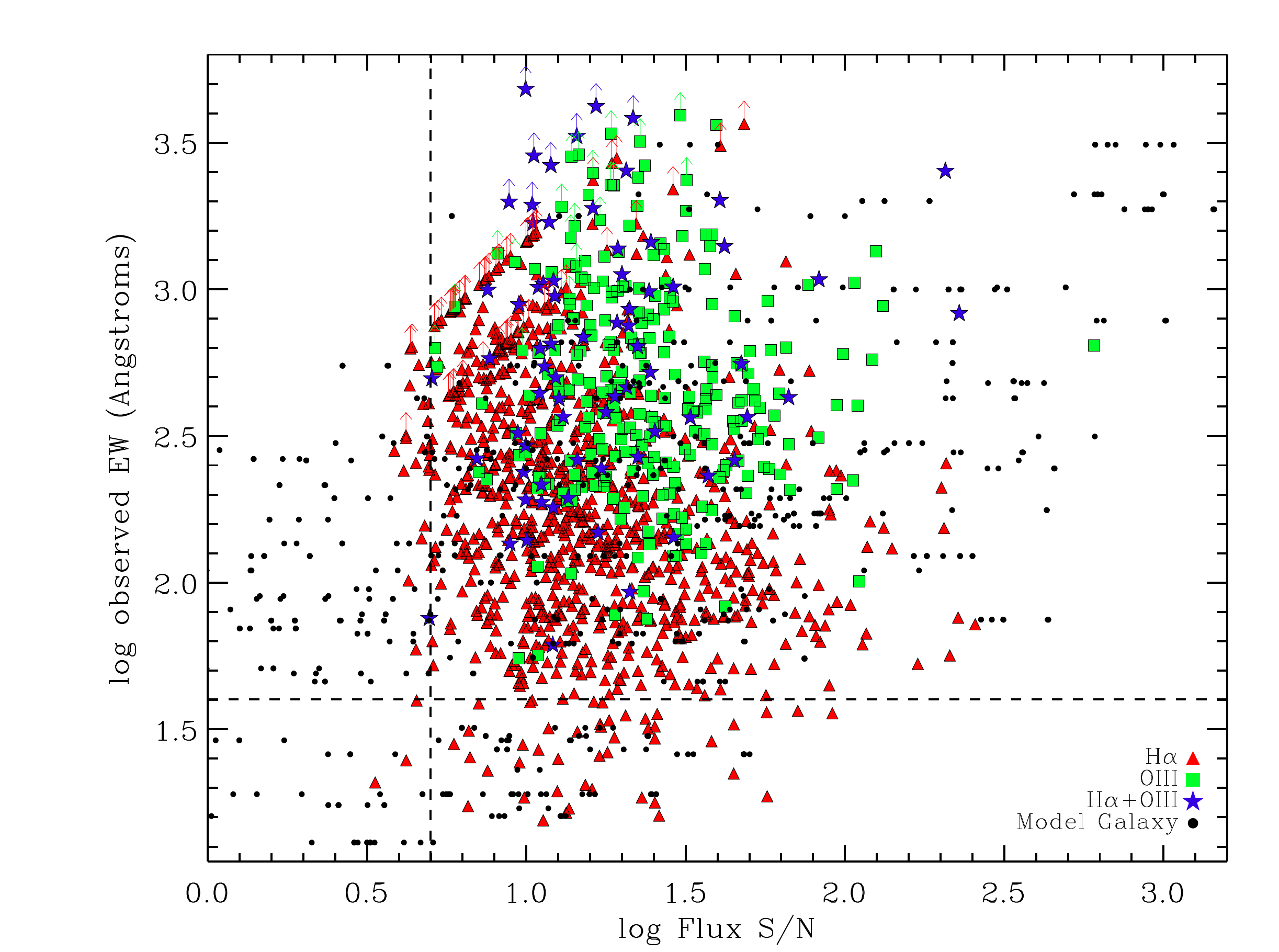}
\caption{Plot of emission line flux signal-to-noise versus equivalent
  width for all the identified WISP emission lines. Red triangles are
  objects with \han emission lines, green squares are objects with [O{\sc iii}] emission, and blue stars are objects where both lines are
  detected. Where both \han and [O{\sc iii}] are detected, the
  parameters plotted are sums (EW) or quadratic sums (S/N) of
  the two lines together. Over-plotted as solid black circles are the model emission line
  galaxies used for our completeness simulations. The dashed lines are
  the limits in flux and EW applied to our sample. Emission lines that lie beyond
  those limits have been excluded from the final analysis.} 
\end{center}
\end{figure*}

\begin{figure}
\begin{center}
\includegraphics[width=3.5in]{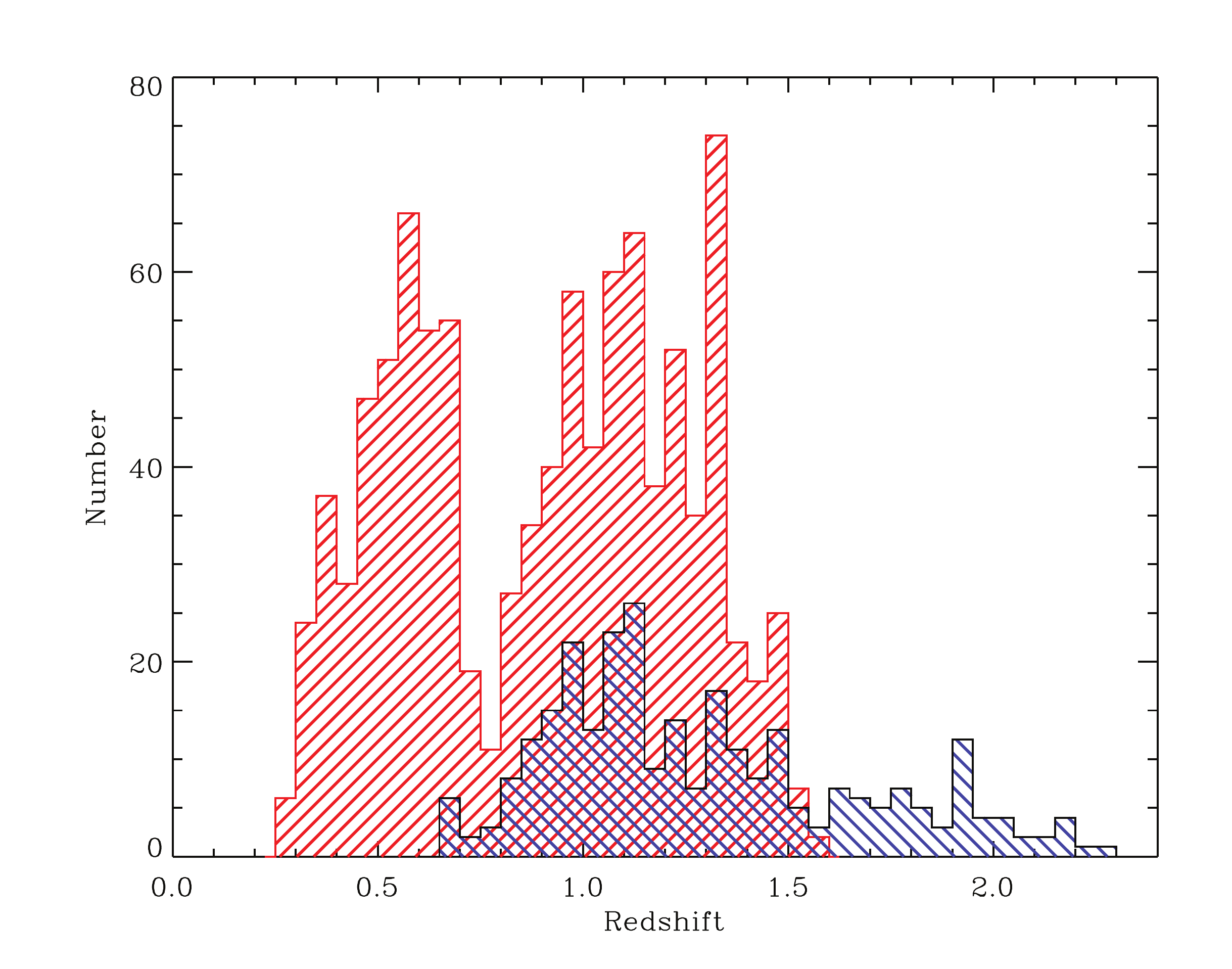}
\caption{Histogram of redshifts for all emission line objects. The
  larger distribution (red) is objects with \han emission, while the smaller
  distribution (blue) is objects with [O{\sc iii}] emission. Objects
  with both \han and [O{\sc iii}] emission are included within both
  distributions and can be approximated as the region where the two
  distributions overlap (there are only 2 $z$$<$1.5 [O{\sc iii}] emitters
  without a \han detection).} 
\end{center}
\label{zdist}
\end{figure}

Single emission lines are assumed to be \ha , except in the few rare
circumstances where the single line is clearly resolved into the [O{\sc
  iii}] 5007+4959 \AA\ doublet. Our simulations indicate that this is a good
assumption and does not result in more than 6\% of our \han lines
being actually misidentified [O{\sc iii}] (see Section \ref{contam}
below). However, this assumption does have a large effect on the recovery of
high redshift (z$>$1.5) [O{\sc iii}] lines, requiring some additional
completeness corrections for those emission lines (see Section \ref{o3completeness}).
    
In total, we extracted 1960 emission lines from the 29 fields down to a
typical flux limit of 3-5$\times$10$^{-17}$ ergs s$^{-1}$ cm$^{-2}$. For an emission line
galaxy to be included in our final analysis we also require an
observed equivalent width greater than 40 \AA\ (log EW = 1.6; see
Section \ref{completeness} below) and a signal-to-noise greater than 5
for at least one of the detected emission lines, creating a final sample size
of 1048 galaxies. Of these, 467 have multiple detected emission lines,
including weaker [O{\sc ii}], \hb , and [S{\sc ii}].
For our final sample of emission line galaxies, we detect \han in 996
galaxies and [O{\sc iii}] in 280 galaxies. The overlap where we found
both \han and [O{\sc iii}] in the same galaxy is 226.

\subsection{Completeness Corrections}
\label{completeness}

Virtually all surveys suffer incompleteness from objects lost as the
strength of their signals approaches the level of the noise. 
A slitless grism survey, like WISP, also suffers significant
incompletenesss due to confusion from nearby
bright sources. In addition to those two fundamental sources of
incompleteness, emission lines can also be lost as part of the extraction
methodology. These sorts of failures include objects missed by the 
automatic line finding routine, objects removed due to redshift
confusion, and objects incorrectly rejected by visual inspection. In
order to derive final completeness corrections to account for all
the forms of incompleteness that effect the grism data, 
we must simulate the entire line extraction process from beginning to end.

We start with the generation of a set of emission line galaxy models, covering the full range
of redshifts ($z$=0.3-2), brightnesses (19-26 F110W mag), spatial extent (0.2-1\arcsec radius),
and equivalent widths (20-2000 \AA) found in our WISP data set of emission line
galaxies. 
We have used two different spectral templates extracted from the
Kinney-Calzetti Atlas in the STSDAS/SYNPHOT library. 
The spectra are from real starburst galaxies observed over the whole
UV and optical range up to 1 \micron. 
We have chosen two spectra with relatively flat continuum in $f_{\nu}$
and different line ratios of \ha/[O{\sc iii}] $\sim$ 4.3 and 0.75. 
The size of the galaxies were then simulated using random values of
the minor and major axis defined as the 
FWHM of light profile in the SExtractor catalog. Throughout this paper
any reference to galaxy size or galaxy radius always refers to
the size of the direct (F110W) image and not to the spatial
extent of the emission lines seen in the grism spectra.
We place each model galaxy randomly into one of our actual
pairs of WISP G102 and G141 grism images using the aXeSIM software \citep{kum07}, before 
extracting it using the same methodology and pipeline as used for extracting all the
real emission line galaxies. This means that first we identify the lines
using the automated software and then two separate team members
visually examine each spectrum to ensure it contains a real emission line.
During the visual inspection we 
mix 15 model galaxies along with another 15 randomly chosen
galaxies, such that
the reviewer could not assume that each spectrum must contain a
visible emission line. Overall we insert 923 model galaxies (74
fields each with typically 10-15 model galaxies) and, including the
random galaxies, extract twice that number. 

We find that the primary determinants of completeness for the emission
line galaxy sample are the observed equivalent width (EW) and the line
flux of the emission line. In practice, the latter translates to the signal-to-noise of
the emission line flux measurement, as the background noise varied
from field to field by a factor of 4.
We also find completeness had some
dependency on the radius (here defined as the major and minor axis
determined by Sextractor averaged in quadrature) 
of the source galaxy, but that this is of secondary importance due to relatively small number of large ($>$
0.75\arcsec ) radius galaxies in the observed sample, approximately
0.5\% (see Section \ref{oiii_ratio} for further discussion of the size
of emission line galaxies). 

In Figure 1, we plot completeness as a function of these three
parameters (line flux S/N, EW, and size), where each bin is a combination of recovery rates for the 
model galaxies. We weight all the model galaxy recovery rates for each
single parameter by the frequency in which
galaxies with the other two parameters appear in our full sample of
actual emission line galaxies. We acknowledge that these observed frequencies of the
different galaxy parameters are not the absolute true frequencies.
However, the implied differential numbers of ``missing'' emission line
galaxies due to incompleteness are not large enough to make a significant impact on the
final derived incompletenesses, so we weight by the observed parameter frequencies in
the interest of simplicity. For instance, if the full sample has
twice as many small EW objects as large EW objects, model galaxies with small EWs
will be given twice the weight when deriving
completeness for S/N or size. 

One immediate result from this simulation is that the recovery rate
drops rapidly from 60\% to $<$20\% as the observed EW (\AA ) of the emission line
becomes less than 40 \AA\ (log EW =1.6; see Figure 1). In the rest
frame of our $z$\ab 1 galaxies, this EW limit corresponds to 20 \AA ,
roughly two thirds that of the average \han
EW for spiral galaxies in the Local Volume \citep{lee07} and
approximately equivalent to the average \han EW for the most massive
$z$=0.8-1.5 galaxies (11$<$log M$_{\sun }$$<$11.5) identified by 3DHST \citep{fum12}.
In the interest of not introducing extremely large ($>$ 5)
and uncertain completeness corrections to the number counts, we apply a cut of
 EW $>$ 40 \AA\ to all of our analysis. After combining this EW cut with our requirement
that all detected emission lines have a signal-to-noise $>$ 5,  the
total number of model galaxies is 553. Of those
model galaxies, we recovered 380, yielding an approximate recovery rate of 69\%. We
note that for this combination of equivalent width and signal-to-noise,
there are effectively no galaxies fainter than 25 magnitudes 
(as measured at F110W-band, although the simulated galaxies are
flat in F$_{\nu }$) in the final sample we use for our analysis.

In Figure 2, we present emission line EW as a function of
signal-to-noise for both observed and model emission-line galaxies. 
One can see that the model galaxies
span the same parameter range as the real galaxies, extending farther
to brighter lines with large EWs and fainter lines with smaller EWs in order to test how
our sample completeness acts at these extremes. One complication is that over a
significant portion of the redshift range ($z$=0.7-1.5), there are
actually two strong emission lines, [O{\sc iii}] and \ha . 
In some cases [O{\sc iii}] is the more powerful line, particularly as we go to
fainter line fluxes (see Section \ref{oiii_ratio} below). We therefore used
model templates both where \han is more powerful and ones where
[O{\sc iii}] is the stronger emission line (\ha /[O{\sc iii}] = 0.75
-- 4.3). We plot the redshift distributions for the detected \han and [O{\sc
  ii}] emitters in Figure \ref{zdist}.  The gap in the \han
distribution around $z$=0.8 is a result of
the low signal-to-noise wavelength range edges of the G102 and G141
grisms where they overlap at  1.1--1.2 $\mu$m. 

One would expect a source with multiple
lines to have a higher completeness than a single line source with the
same line flux and EW. Therefore to determine the likelihood of
identification, we measure the combined signal-to-noise of both
lines, add them quadratically, and use that total signal-to-noise for calculating our
completeness. For the combined EW we use a simple sum of the
two emission lines. We plot these combined values of signal-to-noise and EW as
blue stars in Figure 2, along with the single line cases of \han (red
triangles) and [O{\sc iii}] (green squares). 
In most cases one line is clearly stronger, so
the final signal-to-noise used for the completeness correction can
usually be approximated as the signal-to-noise of just the brightest
emission line.  
We do not use the other, weaker emission lines ([O{\sc ii}], [S{\sc ii}], \hb ,
etc.) for determining completeness, as these lines are never alone and their relative fluxes are
almost always small compared to \han and [O{\sc iii}].

We find that for even the highest signal-to-noise lines (S/N $>$ 100),
we reach only 90-95\% completeness. The main reason for
this is contamination, either from overlapping spectra or large, bright objects in
the direct images used to identify the galaxy counterparts. Either 
the line is lost under an extremely bright spectrum (most fields have 
3-5 moderately bright H$<$19 objects), the object lies in a complicated region where we
are unable to determine which of the nearby, overlapping objects to associate with the emission
line, or the contamination from overlapping sources has caused the automatic line finder to
fail, i.e., Sextractor fails to find the source or the
continuum fit fails badly. Altogether contamination from nearby
sources explains about 75\% of line extraction failures. 
Most nearby source contamination issues would be resolved if data could be
taken at two or more roll angles, such that the spectra do not
overlap, but that is not 
an option for parallel programs.
Another reason a strong line will fall out of our sample is if
the two emission line reviewers can not come to consensus on the
redshift of the object (10\% of the bright extraction failures), which
most commonly occurs where it is unclear whether the line is \han or [O{\sc iii}]. For
instance, when a source is spatially extended the
grism wavelength resolution is effectively lowered, making it
difficult to distinguish between the
[O{\sc iii}] doublet and a single emission line. 
This sort of confusion almost never occurs in G102 (R\ab 210),
suggesting that data taken with a higher spectral resolution would
largely solve this problem as well. The remaining \ab 15\% of bright
emission line failures are mostly miscellaneous extraction errors
by the visual line inspectors. 

\begin{figure}
\begin{center}
\includegraphics[angle=90,width=3.5in]{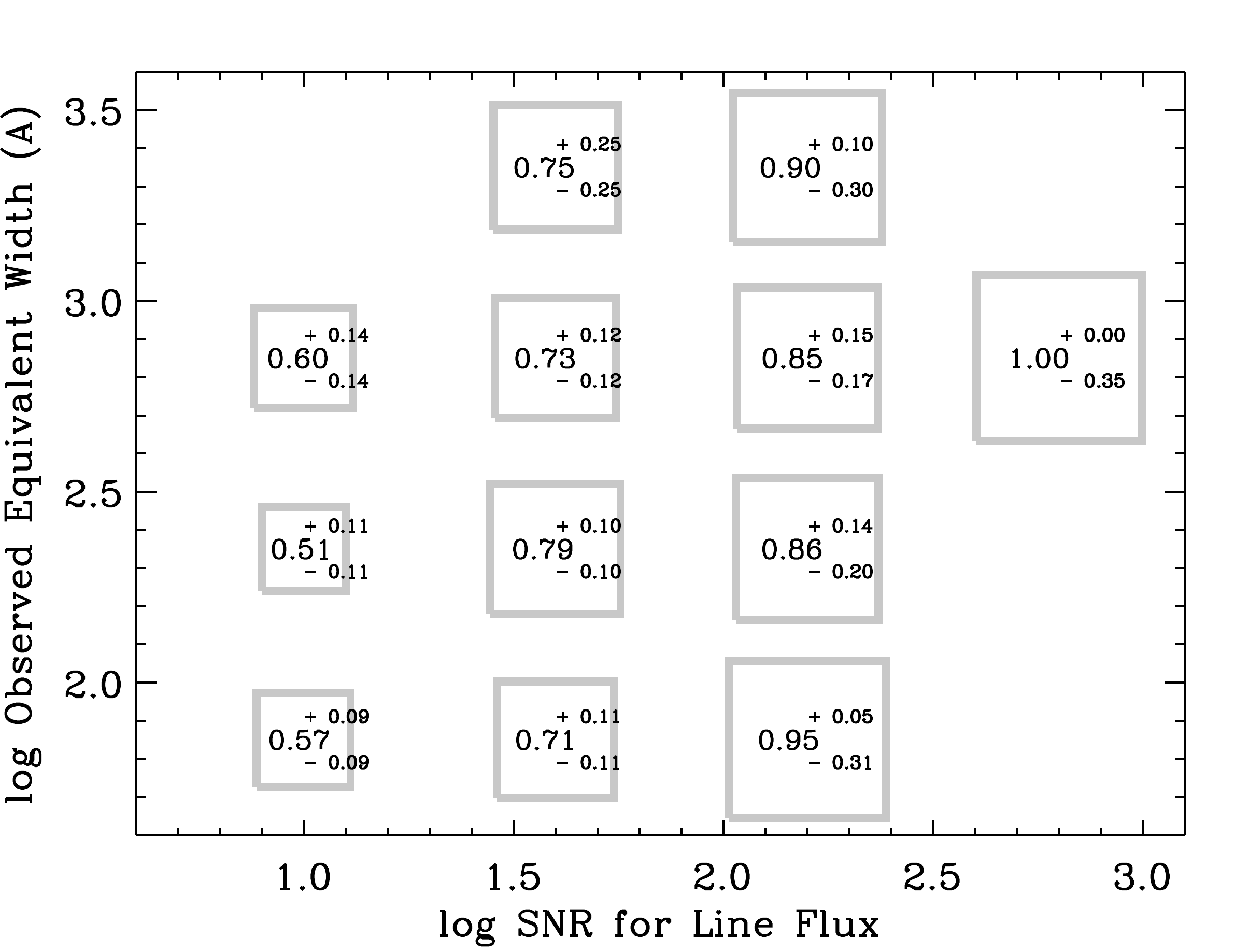}
\caption{Diagram of completeness as a function of both line
flux signal-to-noise and equivalent width. Each box represents one of the two
parameter bins used, with the measured completeness ratio number inside it, including
its error. The larger the square box, 
the smaller the completeness corrections that were applied for those
emission line galaxies.} 
\end{center}
\end{figure}

In Figure 4, we plot the final completeness corrections used for all
of our analysis, derived as a function of both the equivalent width and the
signal-to-noise of the line flux. While the object size
also has an effect on whether an emission line galaxy will be
recovered, we found that accounting for galaxy size has little effect on the
overall completeness corrections, as the largest galaxies make up 
only a tiny percentage of the sample, $<$2\%,
even after accounting for their higher incompleteness. 
Or to put it another way, galaxy size is not very important to the overall completeness
correction, because size completeness corrections grow larger
as the actual number of galaxies is decreasing, unlike EW or line
flux, where the number of objects tends to grow just as their
completeness corrections are significantly increasing.
A three parameter completeness correction is
also not practical  for the required double visual inspection
of each suspected emission line. Each added parameter increases
the human workload geometrically. Instead, for each bin of EW and line flux
signal-to-noise, we weight the input model galaxies by the frequency that
each galaxy size appears in our observed sample list of emission line galaxies. 

\subsubsection{[OIII] Completeness}
\label{o3completeness}

Generally we found that the completeness corrections we derived for
\han can also be applied to [O{\sc iii}] emission lines. While
[O{\sc iii}] is a doublet with a distinctive line profile that one
might expect could be more easily disinguished from artifacts and noise, for
low signal-to-noise and small EWs where the incompleteness becomes
increasing important, [O{\sc iii}] becomes indistiguishable from a single
emission line. 

This assumption that \han and [O{\sc iii}] completeness can be treated
the same breaks down for [O{\sc iii}] at redshifts of
$z>$1.5. This is a result of the attribution of all single line
emitters to \ha . Below $z$=1.5, \han remains in the wavelength range
that is detectable by the WFC3 grisms and in the vast majority of cases if
we detect an [O{\sc iii}] emission line, the \han emission line will
be detectable as well. This is because even down to our faintest
emission line detections, the \ha /[O{\sc iii}] ratio is usually
greater than 1 and rarely less than 0.4 (see Section
\ref{oiii_ratio}). Above $z$=1.5, the \han emission line is
redshifted out of our detectable wavelength range, so we lose the
ability to confirm [O{\sc iii}] with the usually brighter \han line. If the
[O{\sc iii}] line is bright enough we can still identify it from its
doublet line profile, but the faint [O{\sc iii}]  lines will all be identified as
lower redshift \han lines by default. 

While this misidentification of high-z [O{\sc iii}] produces only a small
contamination of \han (see Section \ref{contam}), our simulations show
we lose nearly 50\% of our high-z [O{\sc iii}] emission lines because
of our inability to resolve the doublet. More
specifically, the fainter the emission lines the
larger the incompleteness. We have therefore added an additional
completeness correction to our $z$$>$1.5 [O{\sc iii}] emission lines of
2.0 for 0.6$<$log S/N$<$1.2 and 1.6 for 1.2$<$log S/N$<$1.8. Higher
signal-to-noise [O{\sc iii}] emission lines require no additional
correction.

\subsection{Redshift Accuracy}

\begin{figure}
\includegraphics[angle=90,width=3.5in]{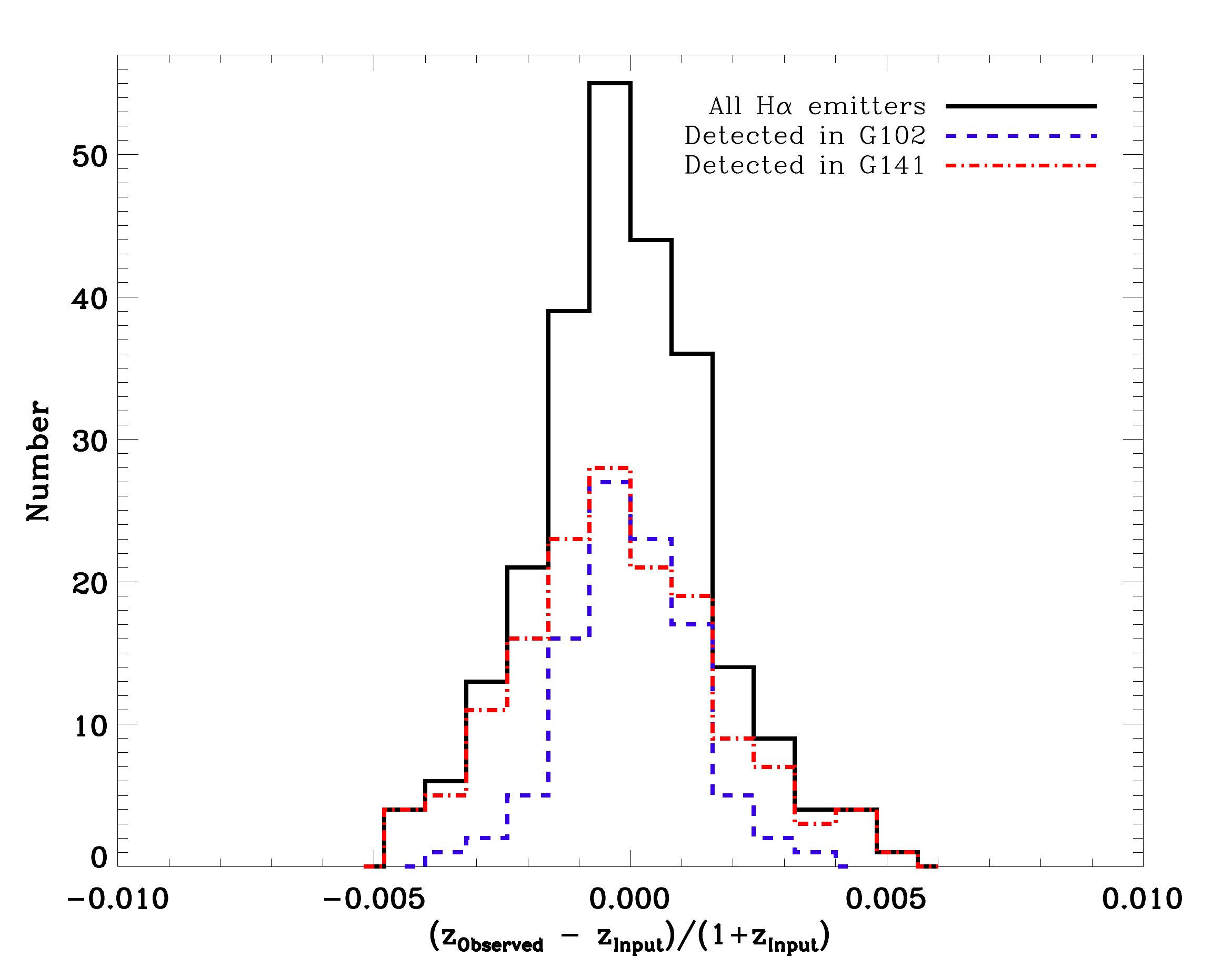}
\caption{Histogram of redshift accuracy for \ha - emitters taken from the simulations, 
where the accuracy is defined as ($z_{observed}$ -
$z_{input}$)/(1+$z_{input}$). The solid black histogram is all \han
emitters, the dashed red histogram is those emitters found only in
the G102 grism, and the dot-dashed blue histogram is the emitters
found only in the G141 grism.} 
\label{redshiftaccuracy}
\end{figure}

Using the simulations we can also get an estimate of the redshift
accuracy which is possible with grism spectroscopy of this wavelength
resolution. Each emission line source is placed at a random position
in the field, which introduces the same uncertainties on the
wavelength solution as seen in real data, including unknown
absolute positions and inexact object sizes. While this is not a measurement of the
absolute wavelength accuracy -- the model spectra are put in and taken 
out with the same software (aXe) and the same assumed wavelength
solution -- it is a reasonable test of the accuracy one can expect
with this pixel sampling, extraction method, and signal-to-noise.  

We removed the emission lines flagged because the reviewers
could not agree on a redshift and then applied our S/N cut of 5 and
EW cut of 40 \AA . We found that going to even higher S/N and larger EW
made almost no difference in the redshift accuracies we derived. For
this analysis we are not interested in the catastrophic errors, which
represent contamination of false lines in our samples rather than a
real reflection of redshift accuracy, so we removed any objects with 
errors in 1+$z$ $>$ 0.5\%. We discuss issues of contamination of false sources
in Section \ref{contam}. We further note that there is
an error in the aXe process of inserting and removing user-generated
spectra that incorrectly shifts the output wavelength calibration by a single pixel,
corresponding to 24 \AA\ in G102 and 46 \AA\ in G141, which we have
removed from all further analysis.

In Figure \ref{redshiftaccuracy}, the histogram of ($z_{observed} - z_{input}$)/(1 +
$z_{input}$) shows that the results are strongly peaked around
zero. The median offset is -0.0002 with a standard deviation
of 0.0017, or roughly 0.2\% accuracy in redshift. Modeling of baryonic
acoustic oscillations suggests that the redshift accuracies required for such
an experiment are more like 0.1\% \citep{wan10}, which is also the
requirement for both the {\it Euclid} and {\it WFIRST} missions. However, it is important
to recall that the WISP redshifts are a combination of two different
near-infrared grisms (G102 and G141) with two different resolutions
(R\ab 210 and R\ab 130). If we split the redshift accuracy
histograms by the grism from which it came, we find the higher
resolution grism reaches accuracies of 0.13\% (as opposed to 0.19\% in
G141), demonstrating that 0.1\% redshift accuracy is possible if the
wavelength resolution R $>$ 200.

It is important to note that the lack of sub-pixel dithering of the parallel observations
prevents any meaningful resampling of pixels along the direction of the
wavelength dispersion. Such resampling would provide a small
improvement in the overall wavelength resolution and would therefore also
potentially increase the redshift accuracies that are possible.

\subsection{False Emission Line Contamination}
\label{contam}

Identifying emission lines in grism spectroscopy
fields without dithering or field rotation is a challenging task. 
Even the well-behaved WFC3 near-infrared arrays
still have cosmic rays, hot pixels, and other
phenomena that can mimic real emission lines. Therefore as we approach
the detection threshold there is a real chance of
contamination of our \han line sample by false emission lines. To investigate this
further we look again to the simulation data. For the completeness
calculation, we track whether an emission line galaxy
is detected and placed into our final emission line list. There
is no requirement that input and output redshifts must match.   
A mismatched redshift indicates contamination. While an emission
line was found, it is clearly not identical to the emission line that
was input. We label all extracted simulated spectra with a difference between input and output (1+$z$)
of more than 1\% as contamination.

Of the 364\footnote[1]{This number is
  slightly less than the 380 simulated emission lines referred to in Section \ref{completeness}, as it
  does not contain 16 weak \han emitters that were found only because of
  their [O{\sc iii}] emission.} simulated \han emission lines that meet all extraction criteria (S/N
$>$ 5, EW $>$ 40 \AA ), 312 are good emission lines found at the input
redshift, a successful extraction rate of 85.7\%. An additional 21
emission lines are at the correct input redshift if we assign them
to be [O{\sc iii}] rather than \ha , for a contamination of \han
emission lines by [O{\sc iii}] of 5.8\%. For the remaining lines there
are no alternative emission lines with which they could be associated,
so they are contamination of non-emission lines to the sample. That
gives us two contamination rates for our sample: 8.5\% contamination 
by false emission lines and 14.3\% total contamination rate of \han emission
lines by the combination of [O{\sc iii}] and false emission lines.
These contamination rates are similar to the 15-30\% contamination
rates found for narrow band \han searches \citep{ly11,sob13}, although
we note that additional information (e.g., galaxy colors, photometric redshifts, and especially
the use of a second narrow band filter) can reduce the narrow band contamination
rate down to 5\% \citep{lee12,sob13}.

This false emission line contamination rate decreases slightly as we increase the
cut-offs of either equivalent width or signal-to-noise of the line
fluxes, but contamination rate never improves to much less than 10\%.
For instance, whether we raise the equivalent width cut-off from 40 \AA\ to 200
\AA\ or raise the the signal-to-noise line flux cut-off from 5 to 20,
the false emission line contamination rate drops by only 3.5\%,
to 10.8\% total contamination.

\begin{figure}
\includegraphics[angle=0,width=3.5in]{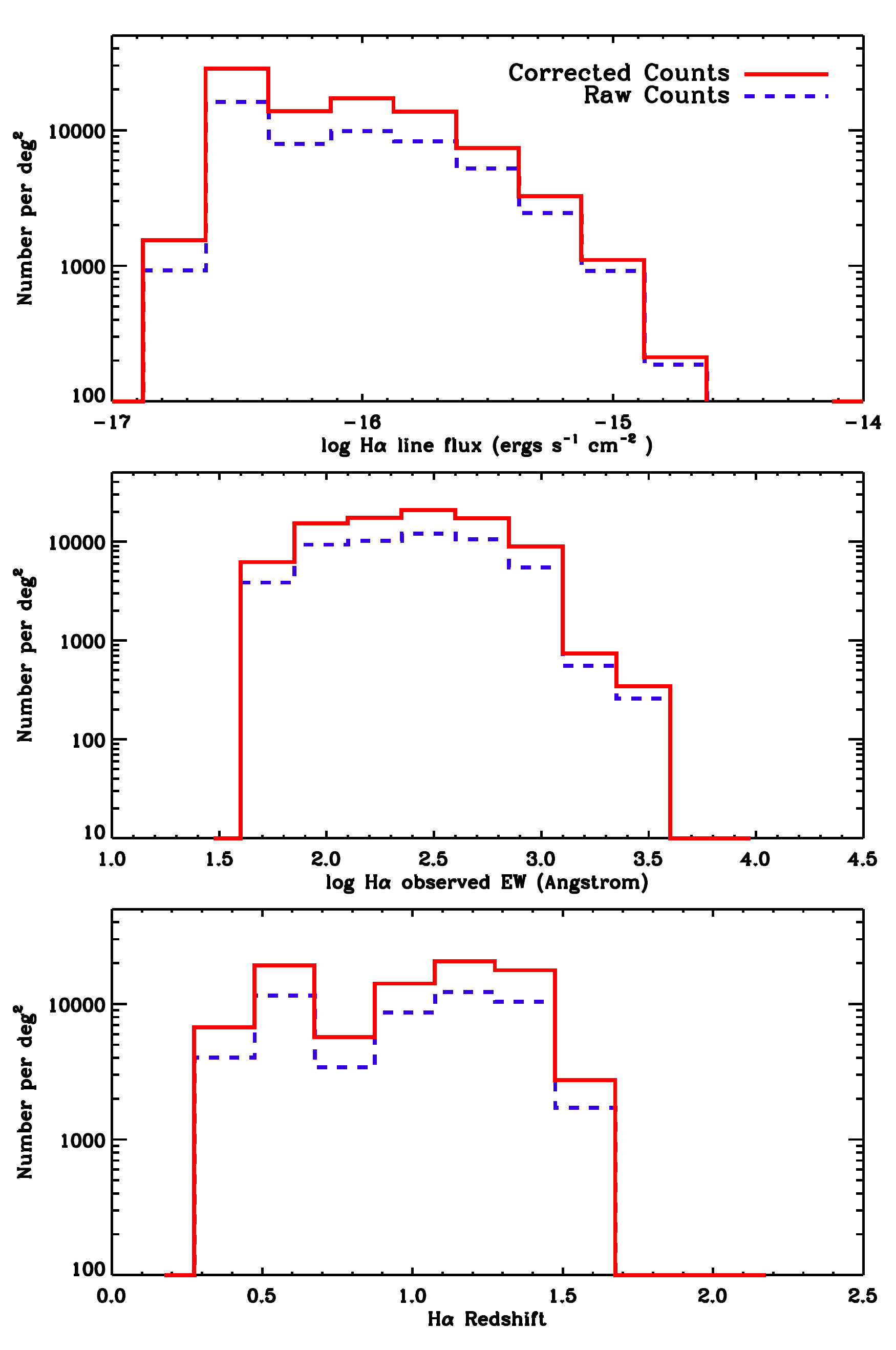}
\caption{Histogram of all detected \han line emitters as a function of
 {\it top)} line flux,  {\it middle)} observed equivalent width, and {\it bottom)}
redshift. The dashed blue lines are the raw, uncorrected counts, while
the solid red lines are the counts after the correction for
completeness. } 
\label{numcounts}
\end{figure}

\begin{figure}
\includegraphics[angle=0,width=3.5in]{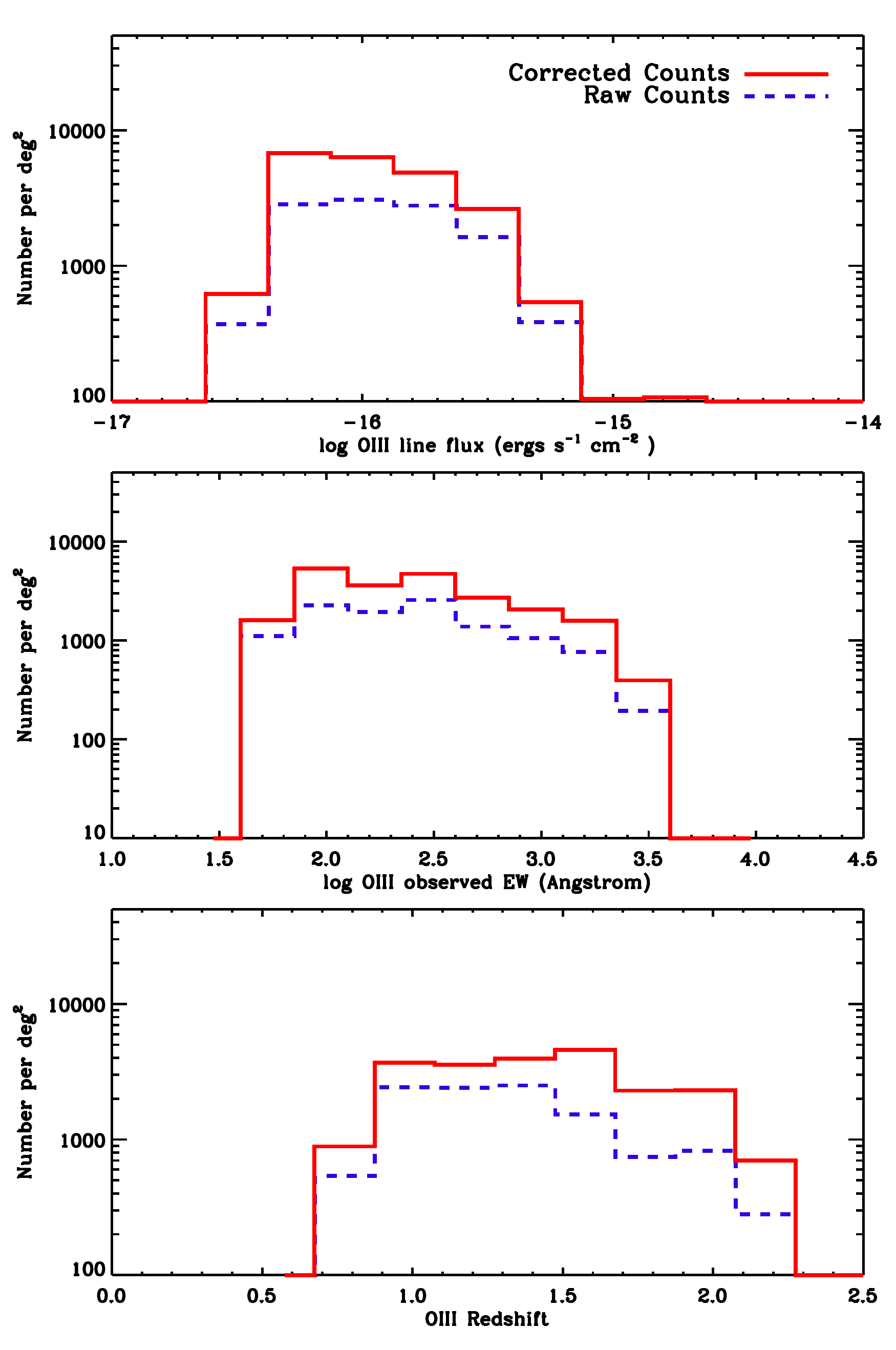}
\caption{Histogram of all detected  [O{\sc III}] line emitters as a function of
 {\it top)} line flux,  {\it middle)} observed equivalent width, and {\it bottom)}
redshift. The dashed blue lines are the raw, uncorrected counts, while
the solid red lines are the counts after the correction for
completeness. We note that the faintest [O{\sc III}] flux bins consist mostly of
secondary, fainter lines that have only been identified and included
because they are companions to brighter \han emission lines. We make no
attempt to correct for this incompleteness, choosing
instead to restrict our analysis to the brighter lines.} 
\label{numcounts_oiii}
\end{figure}

We note that one other potential contaminant is [O{\sc ii}] 3727 \AA .  
[O{\sc ii}] emission is generally faint, with fluxes around 40\% those of
[O{\sc iii}] 5007+4959 \AA . For redshifts where [O{\sc ii}] and [O{\sc iii}] both fall into the
detectable near-infrared wavelength range
(1.3$<$$z$$<$2.3), if we detect [O{\sc ii}], we always detect [O{\sc  iii}] as well. 
There are some rare cases where [O{\sc ii}] 3727 \AA\ can be as much as 30\% brighter than
[O{\sc iii}], but that [O{\sc ii}]/[O{\sc iii}] ratio is not large
enough to make any significant number of [O{\sc iii}]
emitters undetectable. Therefore we do not expect any
contamination from [O{\sc ii}] 3727 \AA\ below a redshift of
$z$$<$2.3. However, the assumption that all single
emission lines are \han could potentially lead us to miss
a population of very high redshift ($z$$>$2.3) [O{\sc ii}]
emitters. These high redshift [O{\sc ii}]-emitters
would have to be very luminous, i.e., more than 5 times the L$_{\star }$
found for O{\sc ii} at $z$=1.47 \citep{ly07}, and should therefore be
quite rare. By $z$=2, the number density of [O{\sc ii}]-emitters
detectable by the WISP survey has dropped below 2$\times$10$^{-4}$ Mpc$^{-3}$. If we allow
for a simple luminosity evolution from
$z$=1.5--2.5 that is consistent with
the observed $z$=2 [O{\sc ii}] number density (a 0.2 magnitude increase in L$^{*}$), then we would not
expect more than a dozen $z$$>$2.3 [O{\sc ii}]-emitters hidden
in our \han sample.

We do not make any separate corrections for these false emission line contamination rates
in our analysis. The completeness corrections we apply are based on
all the extracted emission lines from the simulations,
including the contaminating lines. While the contaminating lines
artificially inflate the number of model emission line galaxies recovered,
which reduces the derived completeness corrections, we expect the same contamination
rates in the actual data. If we used completeness corrections that did not
include contaminating lines, we would incorrectly produce final number counts that
were too large. By leaving the contaminating lines in our completeness
derivation, we produce completeness corrections that simultaneously
account for the effects of both incompleteness and contaminating lines.
To put it another way, the completeness corrections we use are $\sim$ 15\%
smaller than they would otherwise be because of the extra emission
lines we added due to contamination.

\section{Emission Line Number Counts}

The most numerous emission line seen by far is that of \ha , which
makes up two thirds of all those observed. Of the 996 \han emission
lines with EWs $>$ 40 \AA\ and S/N $>$ 5 , 577 (58\%) are identified from a single
line. While identifying \han emission from a single line introduces
the possibility of a misidentification, our simulations (see Section
3.3) demonstrate that this affects no more than 6\% of our emission 
line galaxies. Since virtually all the misidentifications are single
emission lines, this implies that roughly 10\% of our single \han emission
lines are actually not \han at all, but instead misidentified
[O{\sc iii}]. While far from negligible, the level of [O{\sc iii}] emission
line contamination
down to the flux levels probed by this work (3-5$\times$10$^{-17}$ ergs s$^{-1}$ cm$^{-2}$) remains small
enough (10\% of single emission lines, 6\% of all lines) 
that it can be easily accounted for in the statistical analysis.

In Figures \ref{numcounts} and \ref{numcounts_oiii}, we plot the sky
density for \han and [O{\sc iii}] versus flux, equivalent
width, and redshift in both the
original raw counts and those corrected for completeness. Here we can
see that there are significant completeness corrections applied
across the entire range of fluxes. 


\subsection{Emission Line Number Densities}
\label{emissionline}

To derive the number density of the \han emission lines in our
survey, we employed the $V_{MAX}$ method (Felton 1977), in which we
derive the maximum distance for which an emission line of that
absolute luminosity could be detected, depending on the flux limits
of each field, and then translate that to a volume, $V_{MAX}$. This
can be translated into a density:

\begin{equation}
  \delta _{gal} = \frac{4\pi }{\Omega }\sum C(SNR,EW) \frac{1}{V_{MAX}}
\end{equation}

\noindent
where C(SNR,EW) is the completeness correction for each detected
emission line galaxy, as a function of signal-to-noise and equivalent
width. We scale the volume of the entire sphere of the sky by the angular area observed,
4$\pi /\Omega$, to reach the volume density for each object, $\delta _{gal}$.
Then we add up all the $\delta _{gal}$ to get the total volume
densities, which we then use for derivation of the luminosity functions
(see Section 4.3). 

Similarly, we also calculate area number densities, deriving an
$A_{MAX}$ for each source. $A_{MAX}$ is the total angular area from
the survey for which the emission line flux is greater than that of
the 5$\sigma$ line limit of the field. For the
brighter emission lines, $A_{MAX}$ will simply equal the angular area
of the entire survey (29 fields in this case). 
Similar to the $V_{MAX}$ method, once the $A_{MAX}$'s are determined we add up
all the 1/$A_{MAX}$ values to determine the area number density.   
For calculation of both $V_{MAX}$ and $A_{MAX}$, the detector area
used is roughly 85\%  of the
total area of the WFC3 IR chip (4.5 arcmin$^2$), as
objects are lost to the edges where either the 1st order spectrum
falls off the chip (right side) or there are no direct images of
galaxies available (left side).    

The primary surveys of the future near-infrared grism space missions, {\it Euclid} \citep{lau12}
and {\it WFIRST} \citep{gre12,dre12}, will both cover thousands of
square degrees to depths significantly shallower than the WISP
program. The {\it Euclid} wide spectroscopic
survey will cover 15,000 square degrees over a wavelength range of 1.1--2 \m , with a
resolution of R=250 and a 3.5 $\sigma$ line flux depth of 
3$\times$10$^{-16}$ ergs s$^{-1}$ cm$^{-2}$. At
time of publication, the {\it WFIRST} mission is still considering 
several competing designs, but the
primary spectroscopic mission will likely cover at least 2000
square degrees at a resolution of R=600, down to a 7 $\sigma$ line flux depth of 
1$\times$10$^{-16}$ ergs s$^{-1}$ cm$^{-2}$. The wavelength range for
{\it WFIRST} will likely start between 1.3 and 1.5 \m\ and end somewhere
between 2 and 2.4 \m .
While this paper focuses on comparisons and predictions for the
future shallow wide surveys of {\it Euclid} and {\it WFIRST}, it 
should be noted that both missions will certainly contain deeper
surveys done over less area. For instance, the {\it  Euclid}
mission plan contains at least two 20
square degree deep fields with planned depths very similar
to the observations for the WISP program, with line flux limits around
5$\times$10$^{-17}$ ergs s$^{-1}$ cm$^{-2}$.   

\begin{figure}
\includegraphics[angle=90,width=3.5in]{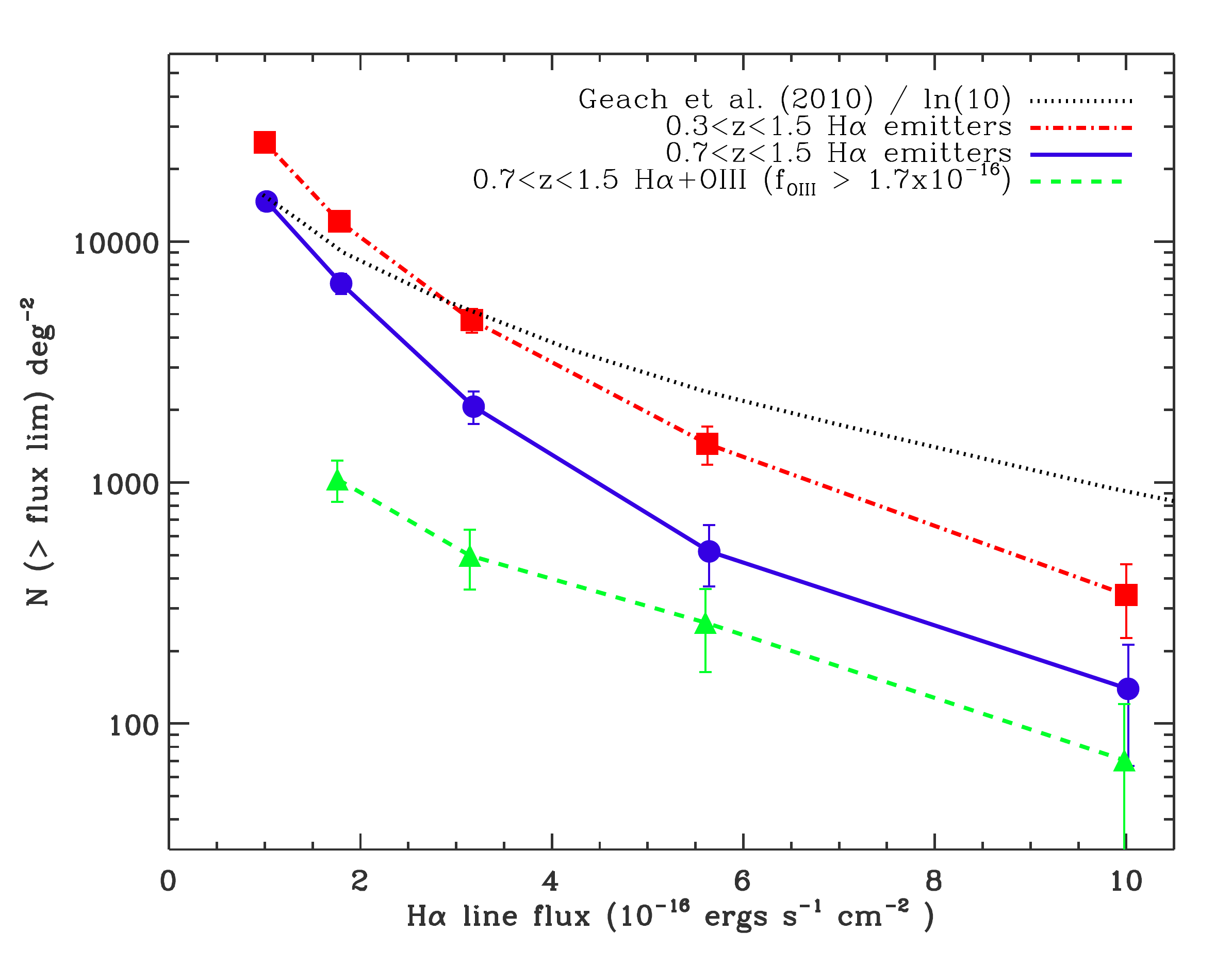}
\caption{Cumulative number count density as a function of \han line
  flux. The red squares and dotted-dashed line are all the 0.3$<$$z$$<$1.5 \han
  emitters, the blue circles and solid line
  are just \han emitters between 0.7$<$$z$$<$1.5, while the green
  triangles and dashed line are
  \han emitters with [O{\sc iii}] line emission greater than
  1.7$\times$10$^{-16}$ ergs s$^{-1}$ cm$^{-2}$ (approximately
  2$\sigma$ for the planned {\it Euclid} line flux limits). The dotted
  black line is the number count density prediction of \cite{gea10}, 
  divided by ln(10). Please note that {\it no} corrections for dust extinction or [N{\sc ii}] contamination have been applied to any
  of the fluxes plotted for cumulative counts in Figures
  \ref{numdensity} -- \ref{numdensity_redshifts_oiii}.} 
\label{numdensity}
\end{figure}

In Figure \ref{numdensity}, we plot
cumulative number counts versus limiting flux for the WISP survey, covering
the primary range of
interest (1-10$\times$10$^{-16}$ ergs s$^{-1}$ cm$^{-2}$) of these
future surveys. Completeness corrections have been applied to all
points and the plotted errors include those derived from our
simulations of the completeness rates (see Figure 4) combined with the 
Poissonian error. There are no corrections to the fluxes for
extinction or the contribution from [N{\sc ii}], which is completely
blended with \han for WISP. 
We present three cumulative number measurements: total \han emitters
(red squares),
0.7$<$z$<$1.5 \han emitters (blue circles), and  0.7$<$z$<$1.5 \han emitters
for which [O{\sc iii}] is also detected (green triangles). At the bright flux
end (10$^{-15}$ ergs s$^{-1}$ cm$^{-2}$) we find a factor of 1.5 times more
\han emission lines at 0.3$<$z$<$0.7 than at 0.7$<$z$<$1.5. As we approach
fainter fluxes the percentage of high redshift objects increases, with
the number of 0.7$<$z$<$1.5 sources exceeding those at 0.3$<$z$<$0.7
around 2$\times$10$^{-16}$ ergs s$^{-1}$ cm$^{-2}$.

\begin{figure}
\includegraphics[angle=90,width=3.5in]{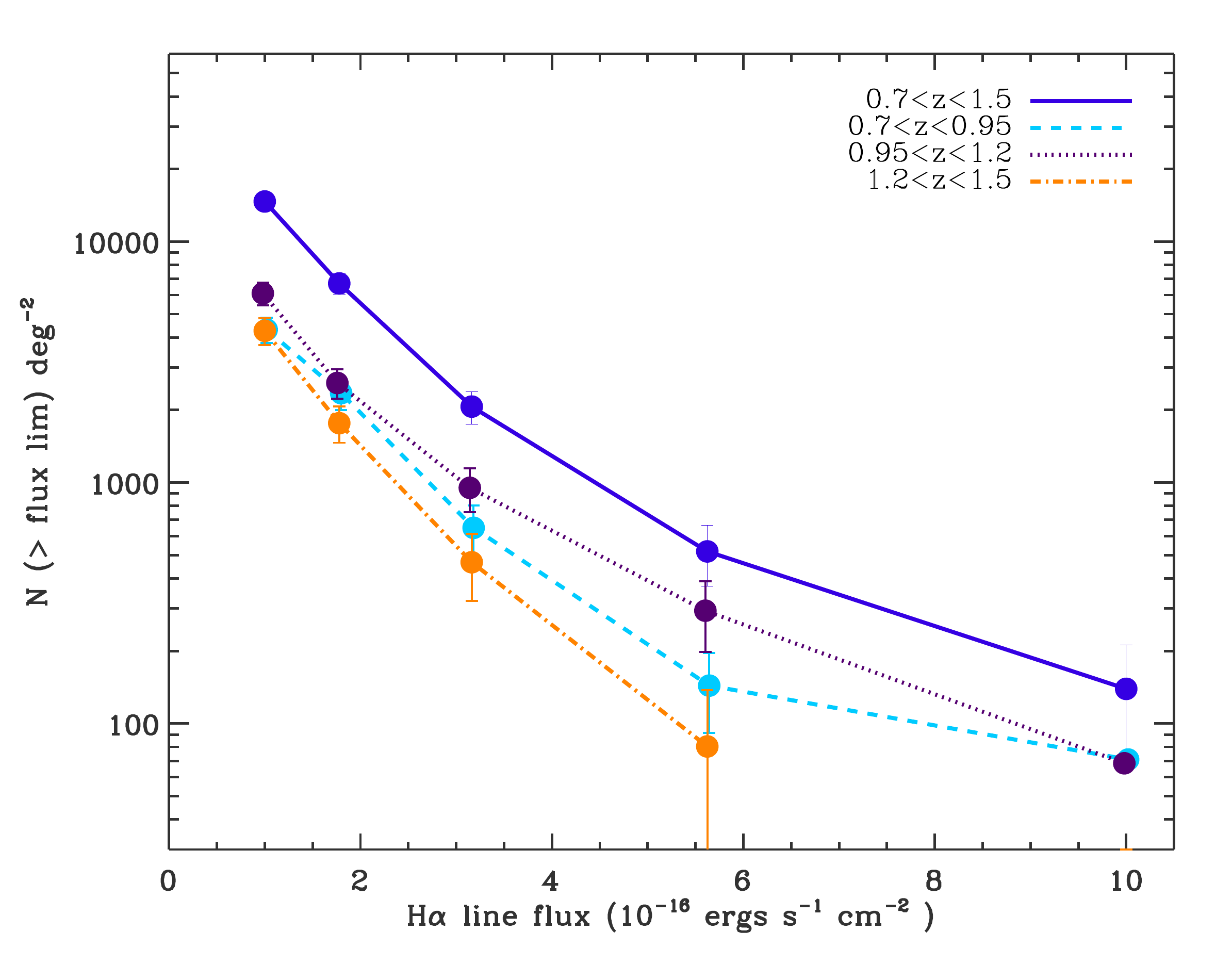}
\caption{Cumulative number count density as a function of \han line
  flux split into different redshift bins from 0.7$<$$z$$<$1.5. The
  solid blue line is all emitters from  0.7$<$$z$$<$1.5 and is unchanged from Figure
  \ref{numdensity}. The dashed light blue line is 0.7$<$$z$$<$0.95, the
  dotted purple line is 0.95$<$$z$$<$1.2, and the dotted-dashed orange
  line is 1.2$<$$z$$<$1.5.} 
\label{numdensity_redshifts}
\end{figure}

\begin{figure}
\includegraphics[angle=90,width=3.5in]{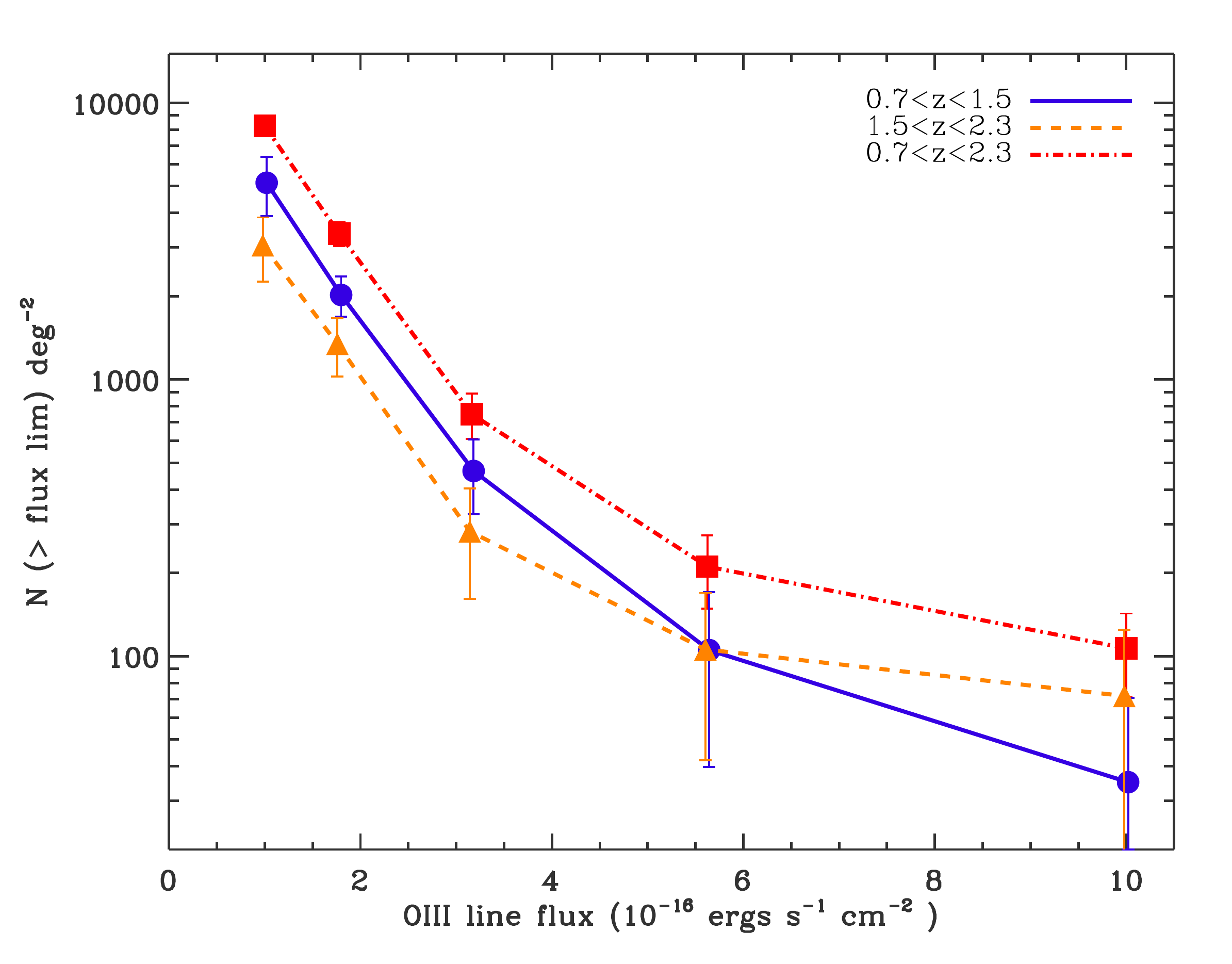}
\caption{Cumulative number count density as a function of [O{\sc III}] line
  flux, split into different redshift bins. The red squares and dotted-dashed
  line is all emitters from  0.7$<$$z$$<$2.3, the blue circles and
  solid line are the number density counts from 0.7$<$$z$$<$1.5, while the
  orange triangles and dashed line are 1.5$<$$z$$<$2.3.} 
\label{numdensity_redshifts_oiii}
\end{figure}

\begin{deluxetable*}{lcccccccc}
\tablecolumns{6} 
\tablewidth{0pc} 
\tablecaption{Emission Line Cumulative Counts\tablenotemark{a}} 
\tablehead{
\colhead{Line Flux \tablenotemark{b}} & \colhead{\han } & \colhead{\han }&
\colhead{\han } & \colhead{\han }&\colhead{\han } & \colhead{\han
 + [O{\sc iii}]\tablenotemark{c}} &\colhead{[O{\sc iii}]}& \colhead{[O{\sc iii}]} \\
 &
\colhead{0.3$<$$z$$<$1.5} & \colhead{0.7$<$$z$$<$1.5}  &
\colhead{0.7$<$$z$$<$0.95} & \colhead{0.95$<$$z$$<$1.2} &
\colhead{1.2$<$$z$$<$1.5} &  \colhead{0.7$<$$z$$<$1.5} &
\colhead{0.7$<$$z$$<$1.5}  & \colhead{1.5$<$$z$$<$2.3} }
\startdata
1.0 & 25806 & 14673 &  4311 & 6098 & 4263 & n/a & 8241 & 3056 \\
1.8 & 12127 & 6703 & 2350 & 2589 & 1764 & 1033 & 3366 & 1345 \\
3.2 & 4721 & 2066 & 649 & 951 & 467 & 498 & 749 & 282 \\
5.6 & 1447 & 517 & 144 & 294 & 80 & 262 & 211 &  106 \\
10.0 & 342 & 139 & 71 & 68 & 0 & 70 & 107 & 72
\enddata
\tablenotetext{a}{All counts are per deg$^{2}$. Fluxes have {\it not} been
  corrected for [N{\sc ii}] contamination.}
\tablenotetext{b}{10$^{-16}$ ergs s$^{-1}$ cm$^{-2}$}
\tablenotetext{c}{Both \han and [O{\sc iii}] required to be greater
  than 1.7$\times$10$^{-16}$ ergs s$^{-1}$ cm$^{-2}$}
\label{emlinecumulative}
\end{deluxetable*}

The WISP survey reaches shorter wavelengths (0.85 \m ) than presently
planned for these future space missions. 
The {\it Euclid} short wavelength limit, 1.1 \m , translates into an \han redshift of $z$=0.7, so we
use the 0.7$<$z$<$1.5 cumulative number counts to predict what these
future space missions will find, noting that the {\it WFIRST} number counts will be
somewhat lower depending on the exact low wavelength cut-off used.
The 1.65 \m\ WISP cut-off is
a shorter long wavelength cut-off than the 2 \m\ or better planned for
these future missions, but that
only means the WISP survey is not sensitive to the highest
redshift \han emitters. Because of
the steep fall-off in number counts with redshift for
the flux limits being planned, it is unlikely the cumulative
\han number counts presented will be much different even with
longer wavelength sensitivity.

We find approximately 2500 (18,500) \han emitters deg$^{-2}$ in the
redshift range 0.7$<$z$<$1.5 down to 3 (1) $\times$
10$^{-16}$ ergs s$^{-1}$ cm$^{-2}$. 
For comparison, we plot
the 0.75$<$z$<$1.9 number counts predicted by the study of \cite{gea10}, which
is mostly based on {\it HST} NICMOS grism and near-IR narrow band
number counts \citep{shi09,gea08,sob09}. We have
decreased the result from the published article by a factor of ln(10) to
account for an error in the published values of $\Phi _{\star }$ used,
a result of improper conversion of $\Phi $(logL)
luminosity functions to the more standard $\Phi $(L) luminosity
functions. We have also shifted the \cite{gea10} predicted counts
to brighter fluxes by a factor of 1.4 to account for the [N{\sc ii}]
contamination which is included in the WISP counts but has been
removed from the counts of \cite{gea10}.
Once we account for these offsets, we find that the NICMOS-derived cumulative
distribution and the WISP distribution approach each other at the
faint end, but that there is an increasingly large disagreement toward
the brighter fluxes, reaching a factor of 6 difference by 10$^{-15}$ ergs s$^{-1}$
cm$^{-2}$. This large disparity at the bright flux end is likely a
result of the NICMOS luminosity functions \citep{yan99,shi09} used to
normalize the \cite{gea10} number count models. The NICMOS luminosity functions contain
significantly more high luminosity ($>$3$\times$10$^{42}$ ergs
s$^{-1}$) \han line emitters than we see for WISP (see Section \ref{lumfunction}).

We are in better agreement with the emission line galaxy count predictions of \cite{wan13},
who also finds a number density of roughly 1000 deg$^{-2}$ down to a \han flux of
4$\times$10$^{-16}$ ergs s$^{-1}$ cm$^{-2}$. Note this flux limit is higher than
the approximately 3$\times$10$^{-16}$ ergs s$^{-1}$ cm$^{-2}$ flux
limit given in \cite{wan13}, but to properly compare the two we had to
add the [N{\sc ii}] contribution back into their \han fluxes. The
\cite{wan13} predictions are based on the \han luminosity functions
from the narrow band survey of
\cite{sob13}, with which we also largely agree (see Section
\ref{lumfunction} below). The general agreement between these two
different emission line detection methodologies (near-infrared grism and narrow
band filters) suggests that the counts presented here are a
robust measurement of the number of \ha -emitting galaxies.

To predict the number of multiple emission line (\han + [O{\sc iii}]) sources a
future near-infrared slitless spectroscopy space mission could expect to find per square
degree, one must first choose a flux limit down to which [O{\sc iii}]
will also be detectable. This detection limit will vary depending on
survey depth and secondary line reliability requirements. We can start
with the presently planned {\it Euclid} mission 
3.5$\sigma$ detection limit of 3$\times$10$^{-16}$ ergs
s$^{-1}$ cm$^{-2}$. If we make the further assumption that one could
reach a lower significance limit of 2$\sigma$ for a secondary line if one already had a higher
confidence line in hand, then one could detect  [O{\sc iii}] flux down
to 1.7 $\times$10$^{-16}$ ergs s$^{-1}$ cm$^{-2}$.  We apply this
[O{\sc iii}] flux limit to the green cumulative count line in Figure
\ref{numdensity}, producing a prediction for {\it Euclid} for the number density of
emission line galaxies where both \han and [O{\sc
  iii}] will be detected. Down to an \han flux of 
3$\times$10$^{-16}$ ergs s$^{-1}$ cm$^{-2}$ we find that roughly 24\%
of the 0.7$<$z$<$1.5 \ha -emitters are also detected in [O{\sc iii}], or
600 \han + [O{\sc iii}] emitters per square degree.
A higher limit on the  [O{\sc iii}] flux will obviously exclude more multiple emission line objects from the
final counts. For instance, if we require the [O{\sc iii}] and \han
line flux to both be greater than 3$\times$10$^{-16}$ ergs
s$^{-1}$ cm$^{-2}$, then the density of \han + [O{\sc iii}] emitters drops
by half, down to around 280 per square degree.

With 314 detected (before completeness correction) \han emission lines at z$>$0.7 with fluxes above 10$^{-16}$ ergs
s$^{-1}$ cm$^{-2}$, we have enough emission lines to break down the
cumulative number counts into even smaller redshift bins in Figure
\ref{numdensity_redshifts}. This more detailed redshift breakdown
allows us to examine the evolution of the number counts that will be available to
these future near-infrared slitless spectrographs in space. At a flux of 10$^{-15}$ ergs s$^{-1}$
cm$^{-2}$ the 0.7$<$z$<$1.5 \han emission line galaxies
are split roughly equally between 0.7$<$$z$$<$0.95 and 0.95$<$$z$$<$1.2. 
However, by 6$\times$10$^{-16}$ ergs s$^{-1}$ cm$^{-2}$, the emitters from the 0.95$<z<$1.2 redshift range outnumber those
from 0.7$<$$z$$<$0.95 by a factors of 1.5-2 and remain more numerous
down to fainter fluxes. Our highest redshift bin, 1.2$<$$z$$<$1.5, has no
emission lines as bright as 10$^{-15}$ ergs s$^{-1}$ cm$^{-2}$ and
is not a significant percentage of the emission line
sample until fluxes less than $\sim$3$\times$10$^{-16}$ ergs s$^{-1}$ cm$^{-2}$.


We can also look at the number densities of the [O{\sc iii}] emission
lines on their own in Figure \ref{numdensity_redshifts_oiii}. Over the same range of line fluxes, the number counts of
[O{\sc iii}] emitters initially rise much more slowly than the
numbers of \ha -emitters, before rising quickly for fluxes fainter
than 3$\times$10$^{-16}$ ergs s$^{-1}$ cm$^{-2}$. 
At 10$^{-15}$ ergs s$^{-1}$ cm$^{-2}$, there are 3
times more \han than [O{\sc iii}] emission lines, but by 
6$\times$10$^{-16}$ ergs s$^{-1}$ cm$^{-2}$ that number density difference has
increased to a factor of 7. Then, thanks to the steep rise of the fainter
[O{\sc iii}]-emitters, this number density disparity rapidly shrinks
and returns to a factor of 3 for line fluxes of 10$^{-16}$ ergs
s$^{-1}$ cm$^{-2}$. This steep rise in the relative [O{\sc iii}] number density
can also be seen in the decreasing H$\alpha$/[O{\sc III}] ratio we
measure for fainter line fluxes (see Section \ref{oiii_ratio} below).

We find roughly 1.5 times as many 0.7$<$$z$$<$1.5 [O{\sc iii}] emitters,
where \han can also be identified in the WISP data, as [O{\sc iii}]
emitters found at the higher 1.5$<$$z$$<$2.3
redshifts. This is after accounting for the extra incompleteness [O{\sc
  iii}]  suffers at the higher redshift (see Section
\ref{o3completeness}). In the actual raw counts we find 3 times as
many 0.7$<$$z$$<$1.5 [O{\sc iii}] emitters. 
This number difference between redshifts remains fairly
constant over the 1-5$\times$ 10$^{-16}$ ergs s$^{-1}$ cm$^{-2}$ flux
range. In the data we see an unexpected crossover towards the
brightest end, with the highest redshift emitters becoming the most
populous. However, the
number statistics here are quite poor, with only 9 total [O{\sc iii}] emitters
covering the 5-10 $\times$ 10$^{-16}$ ergs s$^{-1}$ cm$^{-2}$ flux range, split into the
two redshift bins. Within the error bars, we find the number counts
are consistent with no significant change in the number ratio between
the high and low redshift [O{\sc iii}] emitters over the entire flux
range. This relatively steady number count ratio across a wide range of
redshifts suggests that the luminosity of the typical [O{\sc iii}]
emitter (L$_{\star}$) is increasing strongly with redshift,
something we examine further when we look at the [O{\sc iii}]
luminosity function in Section \ref{lumfunction}. All the cumulative
count number predictions from Figures \ref{numdensity} --
\ref{numdensity_redshifts_oiii} are also listed in
Table \ref{emlinecumulative}.

\subsection{H$\alpha $/[OIII] Ratio}
\label{oiii_ratio}

Over the near-infrared wavelength range probed by the WISP survey,
0.85--1.65\m , the vast majority of the emission lines observed are
either \han or  [O{\sc III}]. When other emission lines are identified they are
always found at the same time as one of those two lines, largely
because these two emission lines are almost always significantly stronger than the other
available lines. Besides the possibility of a rare [O{\sc ii}] emitter (see Section
\ref{contam} above), our simulations
suggest that the assumption that all single line emitters are \han
mainly results in the misidentification of [O{\sc iii}]
emission. While these failed identifications only effect 6\% of \han
lines, it has a large effect on the high redshift ($z$$>$1.5) 
[O{\sc III}]-emitters, where we lose nearly 50\%  from our final sample.

The future near-infrared grism space missions are likely to suffer similar single
emission line misidentification issues. Increased wavelength
resolution will certainly aid in identification, as a resolved [O{\sc
  iii}] 5007+4959 \AA\ doublet will not be confused with \ha . However,
for grism spectroscopy wavelength resolution is not the only limiting
factor. If the emitting region is large in spatial extent it will
smear out the doublet, effectively lowering the available wavelength 
resolution. The median effective radius (the radius within which one half of the total flux is
contained, measured using the F110W filter) of our sample is only
0.3$\arcsec$. Using a similar set of {\it HST} near-infrared 
filters and grisms, \cite{van13} also found effective half-light radii of \ab
0.35$\arcsec$ (3 kpc) across their sample of $z$=0.5-2 galaxies.
However, future wide-area grism missions will
reach shallower flux depths and therefore should find
larger objects, as there is a strong correlation between line flux and
galaxy size, at least for \ha .

\begin{figure}
\includegraphics[angle=90,width=3.5in]{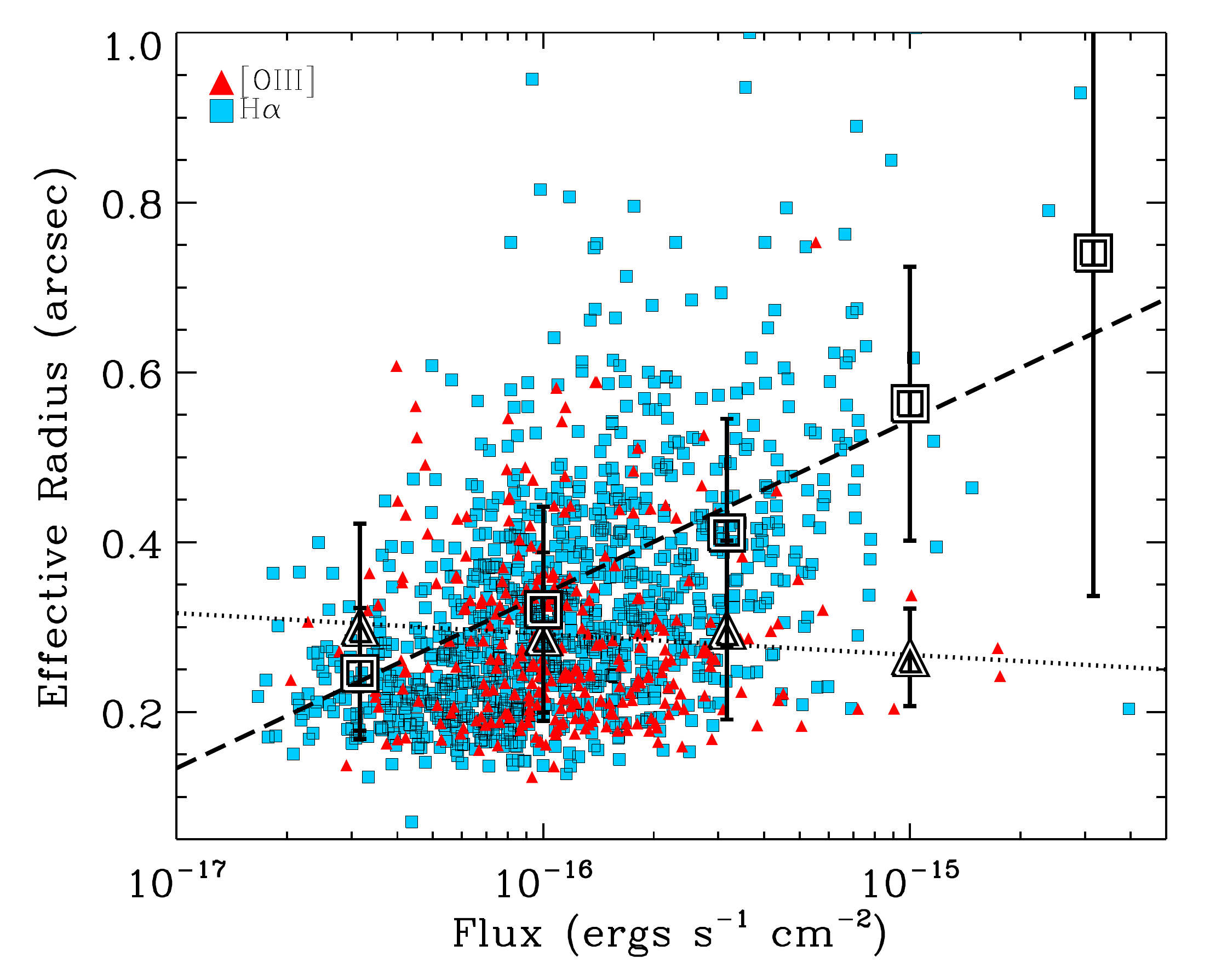}
\caption{Flux versus effective radius for our entire sample of line
  emitters, EW $>$ 40 \AA\ and S/N $>$ 5. \han and \oiii\ emitters are
  plotted as blue squares and red triangles, respectively. The large
  squares and triangles with error bars are the mean effective radii
  for each \han or \oiii\ flux bin. The dashed (\ha ) and dotted
  (\oiii ) lines are the best linear fits to these mean effective radius bins.}
\label{fluxsize}
\end{figure}

In Figure \ref{fluxsize} we plot line flux versus effective
radius for both \han and \oiii\ emission line
galaxies. The large squares (triangles) with error bars are the mean
effective radii for each \han (\oiii ) flux bin, where each input
emission line galaxy has been weighted by the inverse of the
completeness determined for size from Figure 1. Effective radius for \han emitters shows a clear
trend with line flux, r$_{eff}$ = 3.6($\pm $ 1.5) + 0.2($\pm $0.1)$\times$logF,
where the units of r$_{eff}$ are arcseconds and logF is measured in
ergs sec$^{-1}$ cm$^{-2}$. However, there is virtually no trend for \oiii\ emitters at
all, with our best fit being r$_{eff}$ = -0.1($\pm $ 1.1) - 0.025($\pm $0.07)
$\times$logF, essentially consistent with a constant effective radius
over this range of \oiii\ emission line fluxes.     

Down to 3$\times$10$^{-16}$ ergs s$^{-1}$ cm$^{-2}$
and limiting our analysis to redshifts of $z$$>$1.2 (where \ha /[O{\sc
  iii}] confusion becomes an issue at 1.1\m ), we find the radius
of the median object increases 33\% to 0.4$\arcsec$. While the sample
of objects with both high fluxes and high redshifts is small, 13\%
(2 of 15) have radii larger than 0.6$\arcsec$. Fortunately the pixel scale for the 
{\it Euclid} mission will be 0.3$\arcsec$ pixel$^{-1}$, more than twice
that of the IR WFC3, which should greatly mitigate the degradation of 
wavelength resolution caused by extended sources. 
While the exact pixel scale for {\it WFIRST} detectors remains
uncertain, it will likely be close to that of the WFC3 IR channel,
0.13$\arcsec$ pixel$^{-1}$. However, the proposed
{\it WFIRST} wavelength resolution (R=600) is large enough that the
wavelength smearing should not prevent identification of the [O{\sc
  iii}] 5007+4959 \AA\ doublet in most cases.

\begin{figure}
\includegraphics[angle=90,width=3.5in]{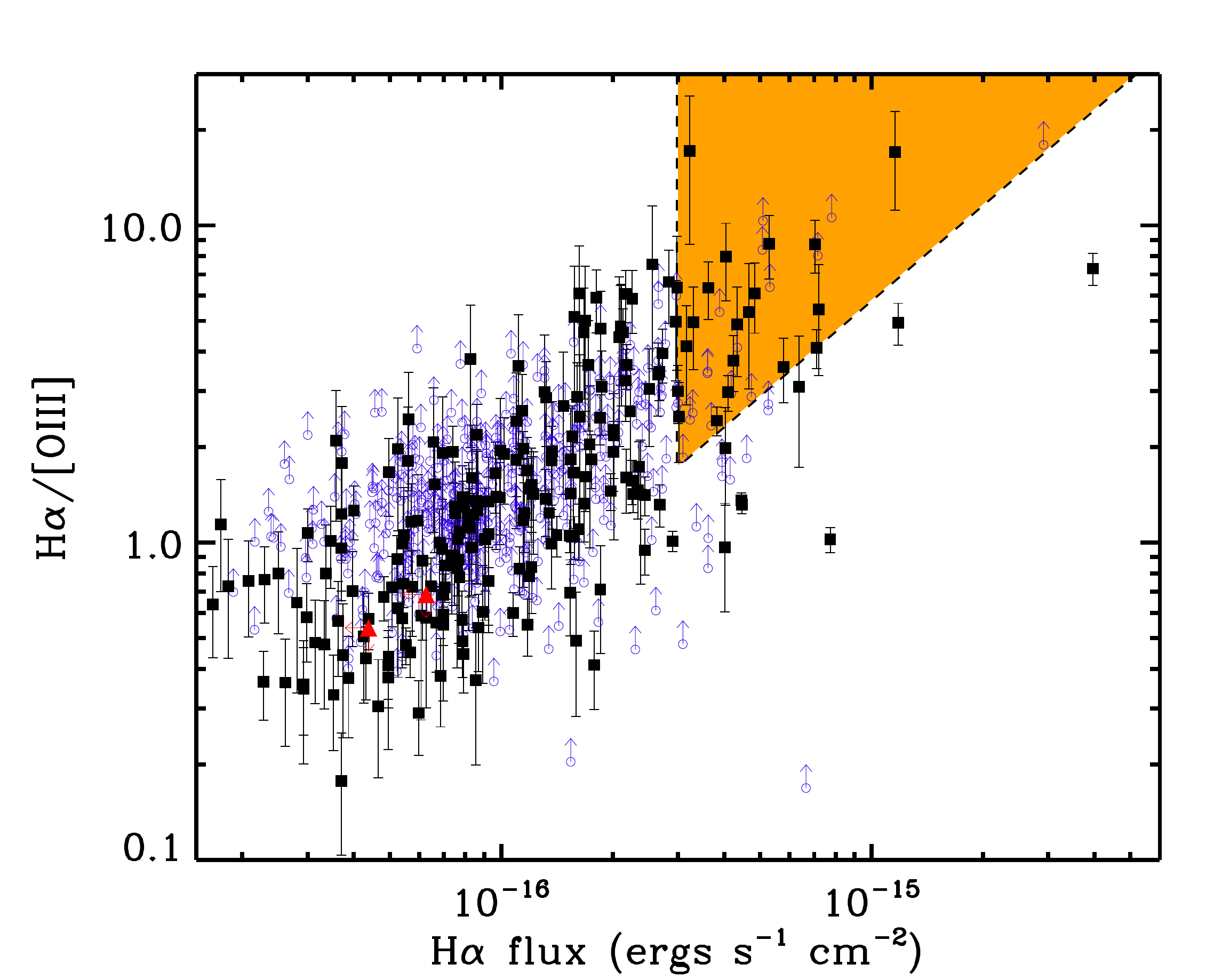}
\caption{Ratio of \han divided by [O{\sc III}] line flux versus \han
  flux for all detected 0.7$<$$z$$<$1.5 line emitters. Solid black squares
  are sources in which both lines are detected. The empty blue circles
  are sources where only \han is detected and are lower limits, while
  the red triangles are the two sources where only [O{\sc III}] has
  been detected, with its two arrows pointing towards the quadrant
  where those points must really lie. The orange wedge represents the area
  where one would expect to find only a single \han emission line,
  assuming an \han detection limit of 3$\times$10$^{-16}$ ergs
s$^{-1}$ cm$^{-2}$ (planned {\it Euclid} survey limit) and a fainter [O{\sc
    III}] detection limit of 1.7$\times$10$^{-16}$ ergs
s$^{-1}$ cm$^{-2}$.}
\label{oiiiratio}
\end{figure}

To investigate the likelihood of single emission line confusion
further, we plot the ratio of \ha /[O{\sc iii}] as a function
of \han flux in Figure \ref{oiiiratio}. We limit our redshift range to that for
which both \han and  [O{\sc iii}] can be detected. Upper limits are 
2$\sigma$, assuming the FWHM of the line for which a limit is being 
derived is the same as for the measured emission line. We find only two  
cases where [O{\sc  iii}]  is detected without \ha . We are able to
identify these two emission lines as [O{\sc  iii}] without secondary
emission lines because they are compact and high enough
signal-to-noise that the distinctive [O{\sc iii}] doublet line profile
is visible.

The upper left part of Figure \ref{oiiiratio}, where the \ha /[O{\sc iii}] ratio is
large and \han fluxes are small, is empty of sources because that is
where the flux limits of the survey prevent sampling. If the high \ha
/[O{\sc iii}] ratios we see at brighter fluxes continue down to
fainter fluxes, then presumably we would find many of the lower
limits moving into that region if we were sensitive enough to detect
such faint [O{\sc iii}] fluxes. The lower right portion of Figure \ref{oiiiratio}, where \ha /[O{\sc iii}] ratio is
small and the \han fluxes are bright, is similarly devoid of objects,
but that region is not excluded by our sensitivity limits. There is a
real scarcity of low \ha /[O{\sc iii}] sources at the bright flux
limits. Excluding three lower limits, we find none of the 58 galaxies
 with \han fluxes greater than 3$\times$10$^{-16}$ ergs s$^{-1}$
 cm$^{-2}$ have ratios of \ha /[O{\sc iii}] $<$ 0.95. 

While the WISP survey does not cover the same wavelength range
as the future space near-infrared grism missions (WISP goes to
shorter wavelengths while the future missions will go longer), it does
give us a rough estimate of
the percentage of single line emitters those missions are likely to find. 
The exact percentage of single line emitters will depend on their final flux
limits and criteria for establishing the reality of an emission
line. For this example we again choose the presently planned
detection limit for {\it Euclid} (3.5$\sigma$ limit of 3$\times$10$^{-16}$ ergs
s$^{-1}$ cm$^{-2}$) and assume a secondary line only needs to be
detected at a 2$\sigma$ significance (1.7 $\times$10$^{-16}$ ergs
s$^{-1}$ cm$^{-2}$). The shaded orange region on Figure \ref{oiiiratio} shows
the region for which only single line emitters would be expected for
these sets of limits. We find that at least 60\% (35/58) of sources
with emission lines brighter than 3$\times$10$^{-16}$ ergs
s$^{-1}$ cm$^{-2}$ will be single line emitters in future
surveys. This single line discovery rate could
potentially be as high as 80\%  if all the lower
limits presented actually lie within the single line discovery region. 

We can also look at a sample where the redshift is $z$$>$1.2, as that
would better reflect the range over which [O{\sc iii}] would be
detectable if the lower wavelength cut-off is the 1.1 \m\
cut-off of {\it Euclid} rather than the 0.85 \m\ of the WISP survey. The number of bright
emitters is much smaller (12) but the percentage of expected single
line emitters, 67--83\%, is virtually the same as when we included the
lower redshifts emitters.  

Our analysis assumes all the single emission lines are \ha , which we know to
be incorrect roughly 10\% of the time. Therefore it is possible that
approximately 3 of the 31 single line emitters in that flux range could be
low \ha /[O{\sc iii}] sources, although if their true redshifts are
greater than $z=1.5$ then their \han fluxes and \ha /[O{\sc iii}]
ratios are not constrained. 
It is important to emphasize that of the 25 sources with
bright fluxes where we detect both \han and [O{\sc iii}], {\it none} of
them have \ha /[O{\sc iii}] $<$ 1. There is no obvious reason why we
would be missing all the galaxies with low \ha /[O{\sc iii}] from
0.7$<$$z$$<$1.5, unless the ratio is so low 
that we lose the ability to detect \ha . However that would require an 
unlikely large gap in the \ha /[O{\sc iii}] ratio, making all galaxies
either strong \han or strong [O{\sc iii}]-emitters, with none in between.

Regardless of the exact quantity of contamination from single line
[O{\sc iii}] emission, clearly \ha /[O{\sc iii}] $<$ 1 sources are much rarer at the
bright fluxes that future near-infrared grism missions will probe than
in our fainter sample. Therefore the assumption that all
single line emitters are \han should produce very few contaminating 
interlopers for these space missions, although further data is
required to better establish exactly what the [O{\sc iii}] contamination rate is.

The trend of increasing \ha /[O{\sc iii}] ratio with \han flux is a result of
the known correlation between the \ha /[O{\sc iii}] ratio and \han
luminosity previously reported by WISP \citep{dom13}. Their analysis
also indicates that dust extinction increases with \han luminosity,
meaning that dust extinction is suppressing [O{\sc iii}] emission more 
at the bright luminosity end. However, the trend of stronger relative
[O{\sc iii}] with fainter \han luminosity is far stronger than their measured
effects of extinction. The dominant determinant of the \ha
/[O{\sc iii}] ratio is most likely the metallicity, as the [O{\sc iii}] 5007 \AA /\hb\
ratio strongly correlates with metallicity in the local universe
\citep[i.e.,][]{lia06}. Metallicity is known to correlate with
mass and luminosity \citep{tre04,kob04}, so it is not surprising that
the brightest \han emitters, which tend to be the most
luminous in the continuum as well, would also have higher \ha /[O{\sc iii}]
ratios.  

Another possible reason for the enhancement of [O{\sc iii}] could be
the presence of AGN, although its effect on the \ha /[O{\sc iii}] ratio
would be at least somewhat mitigated by the similar enhancement of
[N{\sc ii}], which can not be disentangled from \han at the WISP
wavelength resolution. However, line ratios from WISP stacked spectra
are more consistent with the low-metallicity tail
of a star-forming region than that of an AGN \citep{dom13}. 
This agrees with \cite{gar10}, who found $z$=0.84 \ha -selected samples
contain only 5-11\% AGN. \cite{sob13} suggests the AGN fraction
does increase with redshift, but remain only 15\% of the \ha
-emitters above $z$$>$1, with the biggest increases in AGN
contribution at the brightest luminosities. While it is likely that
some of the galaxies with the smallest \ha /[O{\sc iii}] ratios are
AGN, they do not appear to make up enough of the total population of
\ha -emitters to have a significant effect on the overall trend of increasing \ha /[O{\sc
  iii}] ratio with \han flux.

\subsection{Luminosity Functions}
\label{lumfunction}

\begin{figure}
\includegraphics[angle=90,width=3.5in]{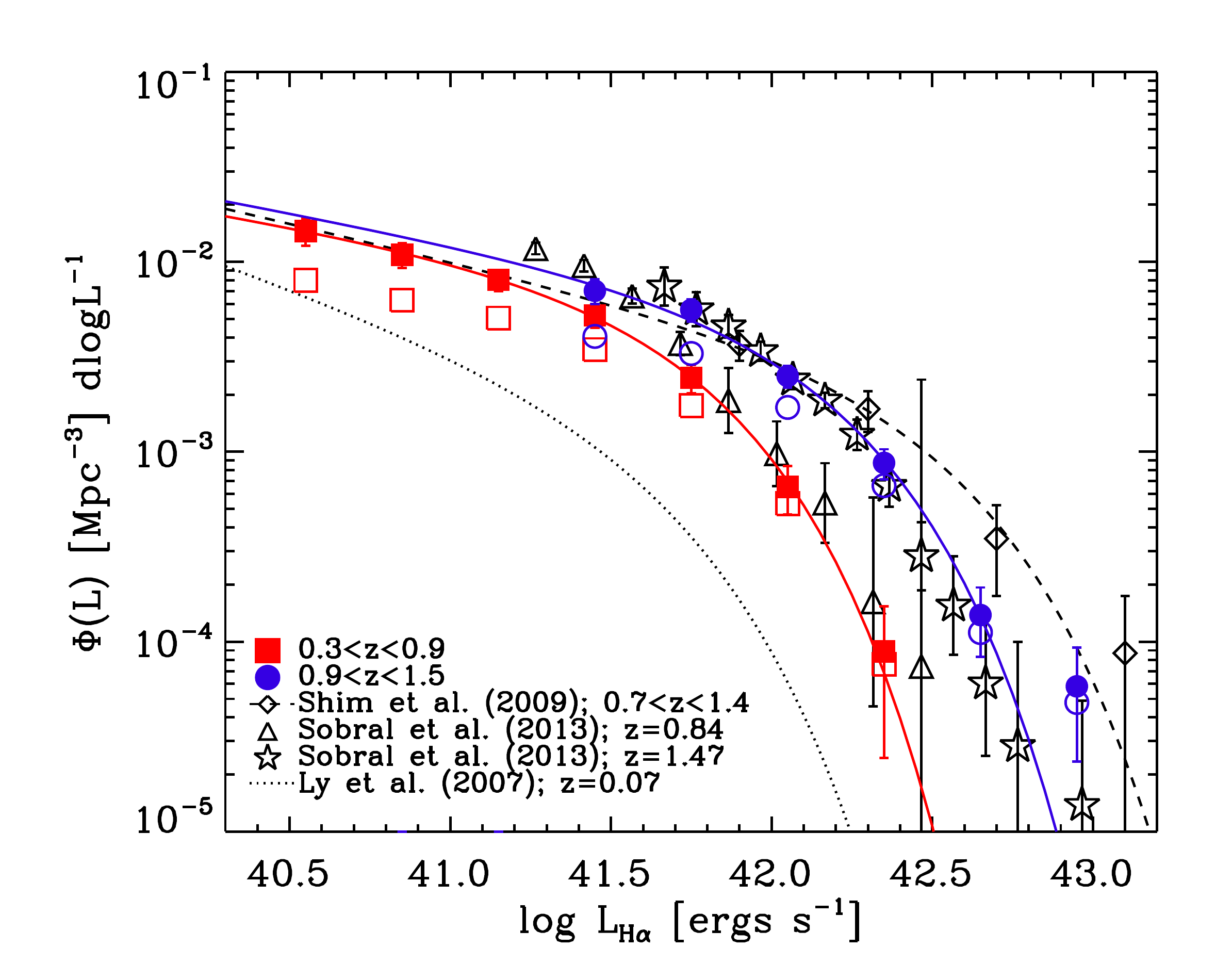}
\caption{\han luminosity function derived from WISP emission line
  sample. The red squares are line emitters from 0.3$<$$z$$<$0.9, while
  the blue circles are line emitters from 0.9$<$$z$$<$1.5. Best-fit
  Schechter functions have been overlaid on top with solid blue and red
  lines. The empty circles and squares are raw counts, while the solid circles and
  squares have been corrected for completeness. The diamonds and dashed
line are the measured points and Schechter fit from the NICMOS grism
study of \cite{shi09} for 0.7$<$$z$$<$1.4. The triangles and stars are from the HiZELS
narrow band study \citep{sob13} at $z$=0.84 and $z$=1.47
respectively. The dotted line is the $z$=0.07 \han luminosity function
\citep{ly07}.} 
\label{halumfun}
\end{figure}

Using the total volume densities derived for each luminosity bin (see
Section 4.1) we are able to plot \han luminosity functions in Figure
\ref{halumfun} and Table \ref{lumfunpoints}.
We have split the \han emission line luminosity function into two
redshift ranges, 0.3$<$$z$$<$0.9 and 0.9$<$$z$$<$1.5. We chose to split the
sample at $z$=0.9 as that divides the detected \han
emission lines by the grism with which they were detected (the dividing line
between G102 and G141 is $\sim$1.2\m ), which also has the added
benefit of producing roughly even sample sizes (436 0.3$<$$z$$<$0.9 versus
517 0.9$<$$z$$<$1.5). We make no correction for dust extinction, but
there is a correction factor of 0.71 applied to the luminosities to
account for [N{\sc ii}] contamination. Please note that before directly comparing
these luminosity functions to the cumulative counts from Section
\ref{emissionline}, the [N{\sc ii}] contamination correction must be
removed.
Our best-fit Schechter parameters are presented in Table \ref{schechter}.

Generally we find that the 0.9$<$$z$$<$1.5  luminosity function is
comparable at the faint luminosity end to luminosity functions
measured by
the previous NICMOS grism studies \citep{yan99,shi09}. However, the WISP number densities drop more quickly as
the \han luminosities become brighter, dropping well below the bright
luminosity number counts from previous NICMOS surveys by
logL=42.5. 
While many of the earlier NICMOS surveys are too small in area for
accurate counts at the bright end of the luminosity function
\citep{yan99,hop00}, the \cite{shi09} luminosity function plotted here
is derived from an area roughly comparable (104 arcmin$^2$) to the WISP data
presented in this work. With more than 139 different fields spread
across the sky, the expected cosmological variance for \cite{shi09} is
also low
\citep[$<$2\%;][]{tre08}, so it can not explain the observed
difference either.
Finally, the redshift sampling difference (0.7$<$$z$$<$1.4 with NICMOS versus
0.9$<$$z$$<$1.5 with WFC3) is not sufficiently different to
effect the luminosity function comparison. The majority of deviation
between the two luminosity functions comes down to a single bin at
logL=42.7, so the observed luminosity function difference may simply
be a result of an unfortunate 2-3 $\sigma$ deviation at that single
point.  


We also find our 0.9$<$$z$$<$1.5 number densities to be consistent with
the $z$=1.47 number densities derived from the narrow
band \han surveys \citep{sob13}.  However, the
redshifts sampled by \cite{sob13} (at $z$=0.84 and 1.47) straddle the sampled range of our high
redshift sample (median redshift of $z$=1.15), so if all the luminosity functions were consistent,
then the WISP luminosity function ought to lie
between the two \cite{sob13} distributions. 
That the number densities at $z$=1.47 and our median $z$=1.15 instead overlap
suggests that the WISP grism survey may be
finding slightly more line emitters than the narrow band survey, although the
absolute differences are very small. Alternatively, the
evolution in the luminosity function between $z$=0.8 and $z$=1.5 could
be occurring mostly at the low end of that redshift range, as
suggested by \cite{gea10}. 
The HiZELS data sample down to very similar EWs (25 \AA ), so 
that is not a likely source of the difference. Neither data sample has
made any attempt to remove AGN, so that can not be a source of difference
either. There is some difference in the extraction apertures used.
HiZELs uses a fixed 2$\arcsec$ aperture at $z$=1.47, while WISP uses a
varying aperture of twice the object size in the direction perpendicular
to the wavelength dispersion (there is effectively no limit to the size
of line emission detectable along the dispersion axis). However, the
majority of objects are compact enough that it seems unlikely either extraction
methodology could be missing much flux.

Our lower redshift
0.3$<$$z$$<$0.9 sample (median $z$=0.58), on the other hand, does fall
below the \cite{sob13} narrow band $z$=0.84 number counts, as well as
the $z$=0.8 narrow band counts (not plotted) from the NEWFIRM \han
Survey \citep{ly11}, although the difference at the bright end of the
luminosity function is within the errors. The more consistent low redshift luminosity densities suggest
that the WISP survey is not generally finding line emitters at a higher rate than narrow
band surveys, but only for its higher redshift bin. A single minor
inconsistency such as this could be partially explained by cosmological
variance, which is expected to be on the order of 15\%
\citep{tre08} for the $z$=1.47 \cite{sob13} narrow band sample.

A significant strength of the WISP survey compared to narrow band
studies is the relatively small contribution of cosmological variance to the overall error
budget. First, the large difference in redshift depth ($\Delta$z =1.2
for WISP versus $\Delta$z = 0.02 for a typical narrow band) means that
despite covering a much smaller area of sky (e.g., 130 arcmin$^2$ versus 1-2
deg$^2$), the WISP survey actually covers a comparable amount of volume.
For instance, a single WISP pointing covers roughly 10$^4$ Mpc$^{3}$
($z$=0.3--1.5), while the $z$=0.84 and $z$=1.47 narrow band samples of \cite{sob13}
cover 10$^4$ Mpc$^{3}$ over 210 and 100 arcmin$^{2}$
respectively. The 29 fields used in this paper survey a volume
of nearly 3$\times$10$^{5}$ Mpc$^{3}$, roughly equivalent to 1 deg$^2$
of the $z$=1.47 narrow band of \cite{sob13}.

Secondly, pencil beam studies like WISP, which cover a small area over a large
redshift range, have significantly lower cosmological variance than
a similar cubical or spherical volume, as the the long, narrow window
must pass through many different environments while a regularly
shaped cube could potentially lie right on top of an extreme
overdensity \citep{tre08}. For objects with a density of 10$^{-3}$ Mpc$^{-3}$, the
cosmic variance is only \ab 12\% for a single WISP pencil beam while it is nearly
60\% for $z$=0.85 and $z$=1.47 narrow beam samples covering the
same volume \citep{tre08}. Of course, the total areas covered by the \cite{sob13}
narrow band surveys are 6-20 times larger (1-2 deg$^2$) from
two separate fields, reducing their associated cosmological variance to 15-20\%. However, as the 
WISP fields are widely separated on the sky, the cosmological variance from field to field is 
uncorrelated, reducing the cosmological variance of the total sample by the
square root of the number of fields. For our example of 10$^{-3}$
objects per Mpc$^{3}$, the contribution of cosmological variance to
the WISP sample presented in this paper is only 2\%. We have included
cosmological variance in our calculations of total errors for the
luminosity functions, although its contribution is essentially negligible
compared to the Poissonian uncertainty and completeness correction
errors.

We note that while a deep pencil beam study like WISP has the advantage of very
little cosmological variance because of the large redshift
range observed, it does so by sacrificing the ability to sample the behavior of
galaxies at more discrete redshifts. In effect, it blurs all the
evolutionary effects that take place within each redshift bin. This
averaging of galaxy evolution is not a significant issue if the
evolution over the sampled redshift range is
taking place at a steady, predictable pace, but could be
misleading if the galaxy behavior changed particularly rapidly or 
became strongly non-linear. Of course, if the evolution becomes too
rapid or unpredictable, narrow band surveys risk completely missing significant behavior that
happens to fall between their sampled redshifts. Because of this, deep pencil beam grism
and narrow band surveys are actually complementary, each a check
on the potential weaknesses of the other.   
   
Regardless of how our sample selection may differ from other surveys,
we can robustly compare how the luminosity function evolves within the WISP
survey as we use the same methodology for selection and completeness
corrections. It appears that most of the evolution from $z=$0.3-1.5
takes place at the bright end of the
luminosity function. Within the errors, the luminosity functions
are converging below logL=41. However, with only two low luminosity bins for the
high redshift sample, this low luminosity covergence is not
definitive. Generally it appears that most of the evolution
in \han from $z$=0.3--1.5 is taking place in  L$_{\star }$, which is
consistent with previous findings \citep{sob13}.  Further
discussion of the \han luminosity functions, their implied star
formation rates, and the evolution of the star formation density, will be
presented in a separate paper (Bunker et al. 2013, in prep).


\begin{deluxetable*}{lccccccc} 
\tablecolumns{8} 
\tablewidth{0pc} 
\tablecaption{\han Luminosity Functions} 
\tablehead{
\colhead{} & \multicolumn{3}{c}{0.3$<$$z$$<$0.9} & \colhead{} &
 \multicolumn{3}{c}{0.9$<$$z$$<$1.5} \\
\cline{2-4} \cline{6-8} \\ 
\colhead{log L\tablenotemark{a}} & \colhead{Number} &
  \colhead{Density} & \colhead{Corrected Density\tablenotemark{b}} & & \colhead{Number} &
  \colhead{Density} & \colhead{Corrected Density\tablenotemark{b}} \\
\colhead{} & \colhead{of Sources} & \colhead{Mpc$^{-3}$
  dlogL$^{-1}$} & 
\colhead{Mpc$^{-3}$ dlogL$^{-1}$} & & \colhead{of Sources} & \colhead{Mpc$^{-3}$
  dlogL$^{-1}$} & 
\colhead{Mpc$^{-3}$ dlogL$^{-1}$}}
\startdata 
      40.55   &    80   &   0.0080  &   0.0145   $\pm\ $0.0024  & & n/a    &          n/a     &                  n/a \\
      40.85    &   97    &  0.0063  &   0.0109  $\pm\  $0.0016 & &  n/a     &          n/a    &                   n/a  \\
      41.15   &    107  &  0.0051  &  0.0080  $\pm\  $0.0010 & &  34  &      0.0087 &   0.0148  $\pm\ $0.0036 \\
      41.45    &   90    &  0.0035  &   0.0052 $\pm\  $0.0007  & & 143      &  0.0041  &   0.0070 $\pm\   $0.0010 \\
      41.75    &   46    &  0.0017  &   0.0025 $\pm\   $0.0004  & & 185 &  0.0033  &   0.0056 $\pm\   $0.0008 \\
      42.05   &    14    &  0.0005  &   0.0007  $\pm\  $0.0002  & & 103 &  0.0017  &   0.0025 $\pm\   $0.0003 \\
      42.35   &    2      &  7.6e-05 &  8.9e-05 $\pm\ $6.5e-05 & & 41 &   0.00066  &   0.00087  $\pm\  $0.00016 \\
      42.65   &    n/a    &          n/a    &                   n/a  & & 7   &  0.00011 & 0.00014 $\pm $ 5.5e-05 \\
      42.95   &    n/a    &          n/a    &                   n/a  & &     3  & 4.8e-05 &  5.8e-05 $\pm\  $ 3.5e-05 \\
      43.25   &    n/a    &          n/a   &                    n/a  & & 1   &  1.6e-05  & 1.6e-05 $ \pm\  $ 1.7e-05
\enddata 
\tablenotetext{a}{Luminosity has been corrected for  [N{\sc ii}]
contamination. No extinction correction has been applied.}
\tablenotetext{b}{Corrected Density is the measured number density with
  completeness corrections applied.}
\label{lumfunpoints}
\end{deluxetable*}

\begin{deluxetable*}{lcccc} 
\tablecolumns{5} 
\tablewidth{0pc} 
\tablecaption{Luminosity Function Best-Fit Schechter Parameters} 
\tablehead{ 
\colhead{Emission Line} & \colhead{Redshift Range} & \colhead{$\Phi _{\star }$}   & 
\colhead{L$_{\star }$}    & \colhead{$\alpha $}
} 
\startdata 
\han & 0.3$<$$z$$<$0.9 & -2.51 $\pm  $   0.11  &      41.72 $\pm  $   0.09
&    -1.27 $\pm  $   0.12 \\

\han & 0.9$<$$z$$<$1.5 & -2.70 $\pm   $  0.12 &      42.18 $\pm  $  0.10 &
-1.43 $\pm  $    0.17 \\

 [O{\sc III}]  &  0.7$<$$z$$<$1.5 & -3.19 $\pm  $   0.09 &     42.34 $\pm  $   0.06 &
-1.40 $\pm  $   0.15 \\

 [O{\sc III}]  &  1.5$<$$z$$<$2.3 &     -3.74 $\pm  $   0.43  &
 42.91 $\pm $ 0.37  &    -1.67 $\pm   $  0.78 \\

[O{\sc III}]  &  0.7$<$$z$$<$1.5 & -3.28 $\pm  $   0.09 &     42.39 $\pm  $   0.08 &
-1.5 (fixed) \\

[O{\sc III}]  &  1.5$<$$z$$<$2.3 & -3.60 $\pm   $  0.14 &     42.83 $\pm   $  0.11 &
-1.5 (fixed) 
\enddata 
\label{schechter}
\end{deluxetable*}

\begin{figure}
\includegraphics[angle=90,width=3.5in]{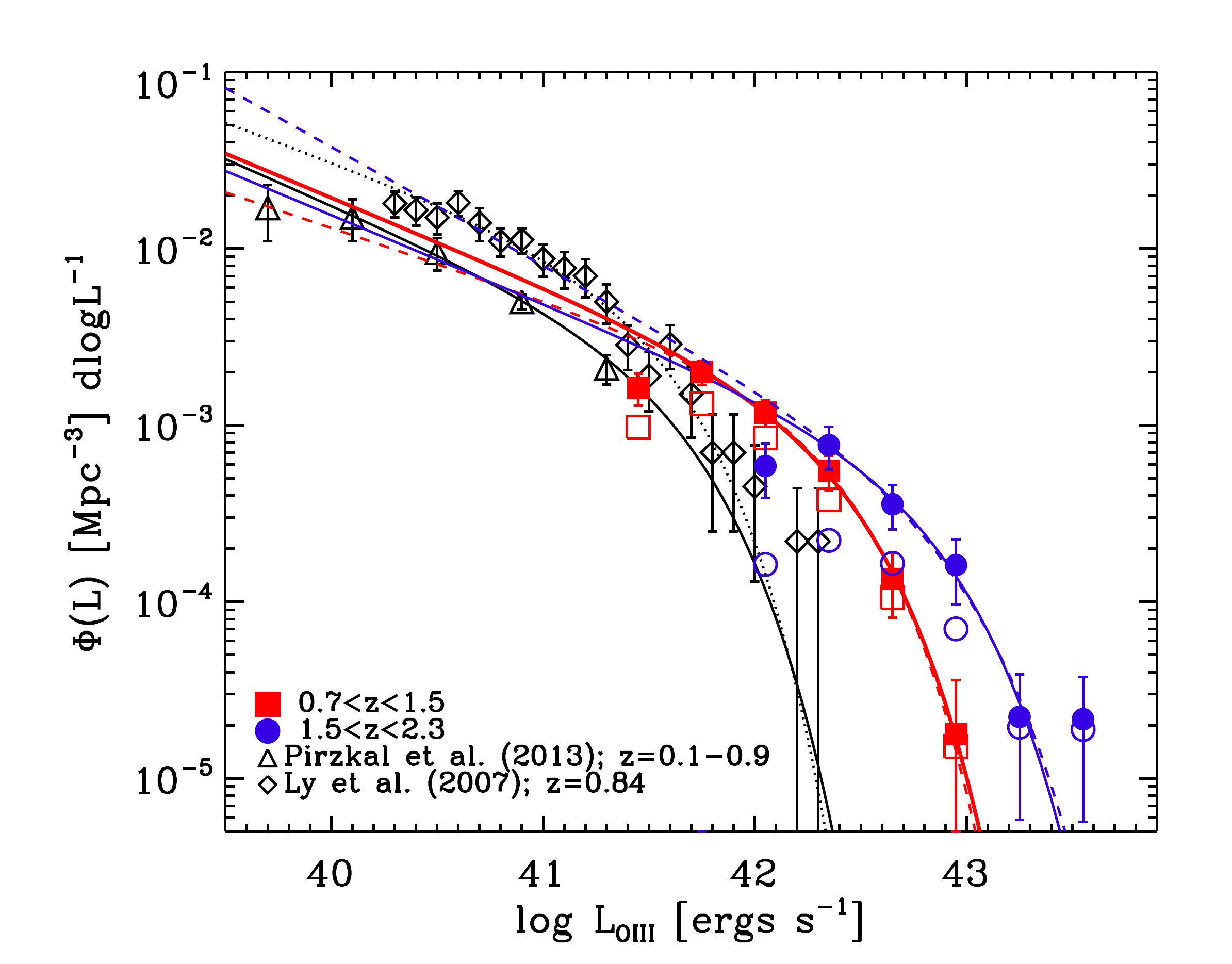}
\caption{[O{\sc iii}] luminosity function derived from WISP emission line
  sample. The red squares are [O{\sc iii}] emitters from 0.7$<$$z$$<$1.5, while
  the blue circles are 1.5$<$$z$$<$2.3. We plot Schecter function fits
  with both a fixed $\alpha$ = -1.5 (solid lines) and one where we allow $\alpha$ to vary
  (dashed lines). The empty  circles and squares are raw counts, while the solid circles and
  squares have been corrected for completeness. The triangles and solid line fit are from the $z$=0.1-0.9 PEARS
Grism Survey \citep{pir13}, while the diamonds and dotted line fit are from the $z$=0.84
narrow band study of \cite{ly07}.} 
\label{o3lumfun}
\end{figure}

\begin{deluxetable*}{lccccccc} 
\tablecolumns{8} 
\tablewidth{0pc} 
\tablecaption{[O{\sc III}] Luminosity Functions} 
\tablehead{
\colhead{} & \multicolumn{3}{c}{0.7$<$$z$$<$1.5} & \colhead{} &
 \multicolumn{3}{c}{1.5$<$$z$$<$2.3} \\
\cline{2-4} \cline{6-8} \\ 
\colhead{log L\tablenotemark{a}} & \colhead{Number} &
  \colhead{Density} & \colhead{Corrected Density\tablenotemark{b}} & & \colhead{Number} &
  \colhead{Density} & \colhead{Corrected Density\tablenotemark{b}} \\
\colhead{} & \colhead{of Sources} & \colhead{Mpc$^{-3}$
  dlogL$^{-1}$} & 
\colhead{Mpc$^{-3}$ dlogL$^{-1}$} & & \colhead{of Sources} & \colhead{Mpc$^{-3}$
  dlogL$^{-1}$} & 
\colhead{Mpc$^{-3}$ dlogL$^{-1}$}}
\startdata 
      41.45    &   36    &  0.00097    &   0.0016  $\pm\  $0.00034  & & n/a  &  n/a  &   n/a \\
      41.75    &   68    &  0.0013      &   0.0020 $\pm\   $0.00032  & &  n/a &  n/a  &   n/a \\
      42.05   &    55    &  0.00085     &   0.0012  $\pm\  $0.00020  & & 11 &  0.00016  &   0.00059 $\pm\   $0.00020 \\
      42.35   &    25      &  0.00038   &  0.00055 $\pm\ $0.00012 & & 20 &   0.00022  &   0.00077  $\pm\  $0.00021 \\
      42.65   &    7      &    0.00011    &  0.00013 $\pm\ $5.3e-05 & & 16  &  0.00017 & 0.00036 $\pm $ 0.00010 \\
      42.95   &    1     &    1.5e-05      &   1.8e-05 $\pm\ $1.8e-05  & &     7  & 7.0e-05 &  0.00016 $\pm\  $ 6.5e-05 \\
      43.25   &    n/a    &          n/a   &                    n/a  & & 2   &  2.0e-05  & 2.2e-05 $ \pm\  $ 1.6e-05 \\
      43.55   &    n/a    &          n/a   &                    n/a  &  &     2 & 1.9e-05 & 2.2e-05 $\pm\ $ 1.6e-05
\enddata 
\tablenotetext{a}{No extinction correction has been applied.}
\tablenotetext{b}{Corrected Density is the measured number density with
  completeness corrections applied.}
\label{oiiifunpoints}
\end{deluxetable*}

We also present the  [O{\sc iii}]  luminosity functions in Figure
\ref{o3lumfun} and Table \ref{oiiifunpoints} along with their accompanying best-fit Schechter parameters
in Table \ref{schechter}. For [O{\sc iii}] we split luminosity functions into two
redshift ranges, 0.7$<$$z$$<$1.5 and 1.5$<$$z$$<$2.3, which are the
highest redshifts for which the [O{\sc iii}] luminosity function has
been determined to date. The lower
redshift bins cover the range over which the
\han line can also be found within the grism spectra. For $z$$>$1.5, \han is redshifted
beyond 1.65\m , leaving [O{\sc iii}] as the most powerful
detectable line in the grism spectrum. The total sample sizes are 
192 [O{\sc III}]-emitters at 0.7$<$$z$$<$1.5 versus 58 at 1.5$<$$z$$<$2.3. 

As remarked in Section \ref{o3completeness}, the [O{\sc iii}] number
densities require higher completeness corrections than for \ha . Even
after application of these additional larger corrections, we still
find that the lowest luminosity bin for each redshift range appears low,
indicating that there is some incompleteness for which we are not
accounting. One possibility is that the number of mis-identified
single line  [O{\sc iii}]-emitters is even larger at the faintest ends
than our simulations indicate. For instance, our simulations
assume \ha /[O{\sc iii}] ratios similar to the emission line sources that we find and
can measure. Therefore, if a population of faint, low \ha /[O{\sc
  iii}] sources exist, we would be underestimating the number of
single line mis-identifications at the lower redshifts (at higher
redshifts the \ha /[O{\sc iii}] ratio does not matter as we can not
see \han at all). Regardless of the reason for the low faintest bin, we
exclude them from the Schechter fits and further analysis. 

Comparing our [O{\sc iii}] luminosity functions to those derived at
lower redshift \citep{pir13,ly07}, we see a strong increase in L$_{\star }$ with redshift
along with decrease in $\Phi _{\star }$. However, our measured
luminosity bins do not constrain $\alpha $ well. If we require
the 0.7$<$$z$$<$1.5 $\alpha $ to be the same as seen at low redshift ($\alpha $=-1.5),
we find only minor changes in $\Phi _{\star }$, with all the evolution
in [O{\sc III}] from $z$=0.7--2.3 taking place in  L$_{\star }$.

\section{Summary and Conclusions}

We present near-infrared emission line number densities, line ratios, and
luminosity functions, based on 29 fields from the WISP survey taken with both
the G102+G141 grism filters. With this dataset we find emission line
galaxies {\it without any pre-selection bias}, over a large continuous epoch
of cosmic time. At least two future space missions, {\it Euclid} and
{\it WFIRST}, will also contain very large near-infrared grism
surveys. The WISP survey is a perfect laboratory to predict what these
future missions can expect to find.

We found the most significant potential issues with identifying emission lines
from near-infrared grism spectroscopy to be:

\begin{itemize}

\item{{\bf Confusion of \han and [O{\sc iii}] due to insufficient
      wavelength resolution,} which can lead to catastrophic redshift
      identification failures. When the signal-to-noise gets low
      enough, one can not necessarily depend on being able to
      distinguish the [O{\sc iii}] doublet from a single emission
      line. This is particularly true in the case where the emission
      line region is spatially extended, which effectively degrades the
      wavelength resolution. This could potentially be a significant issue
      for shallower surveys, as the brighter emission line sources tend to
      also be larger in spatial extent. However, we found this line confusion to
      be almost entirely confined to the lower resolution
      (R$\sim$130) G141 grism and not an issue for the higher
      resolution (R$\sim$210) G102 grism. Additionally our simulations
      found that the higher grism resolution is also needed to achieve the
      0.1\% accuracy in 1+$z$ required for baryonic acoustic
      oscillation experiments. 
      We would therefore strongly
      recommend that future near-IR missions keep to wavelength
      resolutions above R$>$200. }

\item{{\bf Contamination from overlapping spectra from bright sources,}
    which effects even bright, high signal-to-noise spectra. It alters the
    measured continuum, impacts the effectiveness of automatic line finding
    algorithms, makes it difficult to identify the correct host
    galaxy and emission line wavelength, and results in
    emission lines that are lost altogether. 
    Roughly 5\% of all our bright lines ($>$3$\times$10$^{-16}$ ergs
    s$^{-1}$ cm$^{-2}$) are lost to overlapping bright source
    contamination. The use of multiple roll angles would greatly
    minimize these effects, particularly for the brighter, less
    densely-packed sources mostly being targeted by future missions.}

\item{{\bf Lack of dithering,} which not only greatly impacts our
    ability to mitigate cosmic rays, hot and warm pixels, and other
    artifacts that can mimic emission lines, but also prevents us
    from recovering the additional resolution that sub-pixel dithering
    can provide. With grism spectroscopy, this results in not only
    improved spatial resolution, but also improved wavelength
    resolution, which helps to address the insufficient wavelength resolution issue
    already discussed above.}

\end{itemize}

While WISP is sensitive to \han emitters down to $z$=0.3, both {\it
Euclid} and {\it WFIRST}  will likely have low wavelength
cut-offs around 1.1 \m\, making them mostly sensitive to emission line
galaxies at $z$$>$0.7. We find that our cumulative number of \han
galaxies at 0.7$<$$z$$<$1.5 reaches 10,000 deg$^{-2}$ by an \han flux of 2$\times$10$^{-16}$ ergs
s$^{-1}$ cm$^{-2}$. \ha -emitting galaxies with comparable [O{\sc
  iii}] flux are roughly five times less common at these (\ab 1-3 $\times$10$^{-16}$ ergs s$^{-1}$
cm$^{-2}$ ) emission line fluxes.

As emission line fluxes become fainter the \ha /[O{\sc iii}] ratio
becomes smaller, largely a result of the correlation between the \ha
/[O{\sc iii}] ratio and \han luminosity previously reported by WISP
\citep{dom13}. 
While great numbers of \ha /[O{\sc iii}]$<$1 emission
line galaxies can be found around 5$\times$10$^{-17}$ ergs$^{-1}$
cm$^{-2}$, there are effectively none in our sample once the \han flux becomes
greater than 2$\times$10$^{-16}$ ergs$^{-1}$ cm$^{-2}$. These
large \ha /[O{\sc iii}] ratios for the brighter emission lines
suggest that 60-80\% of the \han emission lines found by future space
missions will be single lines. However, that same high \ha /[O{\sc
  iii}] ratio suggests that the likelihood of contamination by lines
besides \han remains
low.

Our \han luminosity function finds a similar
number density of line emitters to that of the NICMOS near-infrared grism
surveys at faint luminosities, but finds significantly fewer numbers of
line emitters (factors of 3-4 less) at the bright luminosity end.
We find our 0.9$<$$z$$<$1.5 number counts to
be consistent with the $z$=1.47 narrow band
\han survey of HiZELS (Sobral et al. 2013), although with our lower median redshift
$z$=1.15, we would have expected slightly lower densities if the
luminosity function is smoothly evolving out to $z$=1.5. On the other
hand, our lower redshift 0.3$<$$z$$<$0.9 \ha -emitter number counts
fall just below both the narrow band counts from HiZELS at $z$=0.84 and
NEWFIRM \han at $z$=0.8.
The absolute difference in expected luminosity density for the higher
redshifts is small and could be partially a result of cosmological
variance of the narrow band sample.

The [O{\sc iii}] emission line probes a higher redshift range than
\han and is therefore not sensitive
to the fainter luminosities required to properly constrain
$\alpha$. If we fix $\alpha$=-1.5, as seen in studies at lower
redshifts, we find that the evolution in the [O{\sc iii}] luminosity function from
$z$=0.7--2.3 is almost entirely in the L$_{\star }$
parameter, similar to what we observe for the $z$=0.3--1.5 evolution
of the \han luminosity function.

\smallskip
We would like to acknowledge the assistance of Chun Ly for his
helpful advice during the production of this paper. The authors 
would also like to acknowledge financial support 
from the grants for HST programs GO-10226,
GO-11696, and GO-12283.


\end{document}